\documentclass[10pt, journal,letterpaper,twocolumn]{IEEEtran}

\usepackage[dvips]{color}
\usepackage{epsf}
\usepackage{times}
\usepackage{cite}
\usepackage{epsfig}
\usepackage{graphicx}
\usepackage{amsmath}
\usepackage{amssymb}
\usepackage{amsthm}
\usepackage{amsxtra}
\usepackage{algorithmic}
\usepackage{float}
\usepackage{amsfonts}
\usepackage{dsfont}
\usepackage{subfigure}
\usepackage{url}

\newcommand{\AlgorithmCaption}[2]{\textbf{Algorithm {#1}} {#2}} 
\newenvironment{boxedalgorithmic}[2] 
  {\noindent\begin{minipage}[t!]{\columnwidth} \hrule \vspace*{0.1cm}\AlgorithmCaption{#1}{#2} \vspace*{0.1cm}\hrule\footnotesize
   \begin{algorithmic}}
  {\end{algorithmic}\vspace*{0.1cm}\hrule
   \end{minipage}}

\newtheorem{Lemma1}{Lemma}
\newtheorem{Lemma2}[Lemma1]{Lemma}
\newtheorem{Lemma3}[Lemma1]{Lemma}

\newtheorem{Thm1}{Theorem}
\newtheorem{Thm2}[Thm1]{Theorem}
\newtheorem{Thm3}[Thm1]{Theorem}

\newtheorem{Corr1}{Corollary}
\newtheorem{Corr2}[Corr1]{Corollary}

\begin{document}

\title{Delay-Constrained Video Transmission: Quality-driven Resource Allocation and Scheduling}
\author{{\large Amin Abdel Khalek,~\IEEEmembership{Student Member,~IEEE}, Constantine Caramanis, ~\IEEEmembership{Member,~IEEE}, \\and Robert W. Heath Jr.,~\IEEEmembership{Fellow,~IEEE}} \\
\thanks{The authors are with the Wireless Networking \& Communications Group in the Department of Electrical and Computer Engineering at UT Austin WNCG, 2501 Speedway Stop C0806, Austin, Texas 78712-1687. Email: \{akhalek,constantine,rheath\}@utexas.edu. This work was partially supported by the Intel-Cisco Video Aware Wireless Networks (VAWN) Program.}\vspace{-0.3cm}}
\maketitle

\begin{abstract}
Real-time video demands quality-of-service (QoS) guarantees such as delay bounds for end-user satisfaction. Furthermore, the tolerable delay varies depending on the use case such as live streaming or two-way video conferencing. Due to the inherently stochastic nature of wireless fading channels, deterministic delay bounds are difficult to guarantee. Instead, we propose providing statistical delay guarantees using the concept of effective capacity. We consider a multiuser setup whereby different users have (possibly different) delay QoS constraints. We derive the resource allocation policy that maximizes the sum video quality and applies to any quality metric with concave rate-quality mapping. We show that the optimal operating point per user is such that the rate-distortion slope is the inverse of the supported video source rate per unit bandwidth, a key metric we refer to as the source spectral efficiency. We also solve the alternative problem of fairness-based resource allocation whereby the objective is to maximize the minimum video quality across users. Finally, we derive user admission and scheduling policies that enable selecting a maximal user subset such that all selected users can meet their statistical delay requirement. Results show that video users with differentiated QoS requirements can achieve similar video quality with vastly different resource requirements. Thus, QoS-aware scheduling and resource allocation enable supporting significantly more users under the same resource constraints.
\end{abstract}

\section{Introduction}

Real-time video transmission requires maintaining stringent delay bounds to ensure a good user experience. The stringency of the delay bound is further dependent on the specific use case. For instance, interactive applications such as video conferencing can only tolerate an end-to-end delay on the order of few hundred milliseconds for a smooth experience whereas with live streaming, the delay constraint can be relaxed to few seconds. In bandwidth-limited networks with shared resources, the bottleneck in the end-to-end delay is queuing. Thus, users with more stringent delay constraints should be allocated more physical resources to boost their service rates, reduce the queuing delay, and thus support their QoS requirement. This provides a strong motivation for re-designing resource allocation taking into account hybrid QoS requirements in the network.

Deterministic delay bounds are hard to guarantee over wireless networks due to changing channel conditions \cite{eff_cap}. Therefore, to provide a realistic and accurate model for quality of service, statistical guarantees are considered as a design guideline by defining constraints in terms of the delay-bound violation probability. The notion of statistical QoS is tied back to the well developed theory of effective bandwidth
\cite{kumar2004communication,eff_band,eff_band2} and its dual concept of effective capacity \cite{eff_cap,wu2006effective,Wu_dissertation}.

Satisfying delay constraints should not come at the expense of maintaining high perceptual quality. Therefore, we derive a resource allocation policy that maximizes a quality-based utility such that all users in the network can achieve their target statistical delay bound. We consider two different utility functions: the sum video quality and the minimum video quality in the network. The only assumption about the quality metric is that the rate-quality mapping is concave which is generally the case because practical video codecs achieve diminishing returns in quality as the source rate increases. For the general case where the set of users in the network cannot all be served, we solve the problem of selecting a maximal subset of users to schedule such that each scheduled user can meet their target QoS requirement.

\subsection{Contributions}

The central contribution of the paper is that it partitions the wireless channel resources across real-time video users with hybrid QoS requirements to maximize a video quality-based utility function. Previous work on delay-constrained video transmission addresses maximizing rate or throughput \cite{tang_qos,vassaki2012effective,ma2012power}, minimizing energy consumption \cite{Khalek_TMC,tang2008cross}, or minimizing resource utilization \cite{stat_qos_uni_multi,tang2008cross}, all of which are not directly relevant metrics for the end video quality. Furthermore, a significant body of literature \cite{tang_qos,vassaki2012effective,ma2012power,Khalek_TMC,tang2008cross,stat_qos_uni_multi} is devoted to point-to-point transmission with resource allocation across different time slots or for different video layers, thus not taking into account hybrid QoS requirements across different users. Previous work which directly optimizes video quality utility functions across different users is either focused on stored video in which content can be buffered ahead opportunistically and large delays can be tolerated \cite{joseph_variability,joseph2012resource} or considers real-time video with deterministic delay constraints \cite{huang2008joint}, thus not being applicable to communication over fast fading channels. Specifically, this paper addresses these issues by answering the following three questions:

\begin{enumerate}
\item \textbf{Resource Allocation:} Given a set of users with specified delay bounds, target violation probabilities, and rate distortion characteristics, how should resources be allocated across the users and how should the source rates be adapted to maximize a video quality-based utility function?
\item \textbf{Scheduling:} If not all users can meet their delay constraint simultaneously, what is the subset of users with the largest cardinality such that all scheduled users can meet their statistical delay bound?
\item \textbf{User Admission:} Given a network in operation and a new video user requesting a session, what is the user admission criterion such that the user can meets the QoS requirement without jeopardizing the QoS of existing users?
\end{enumerate}

In what follows, we summarize the major paper contributions on these three fronts.

\subsubsection{Quality-driven resource allocation and rate adaptation}

Considering a network with multiple video users with possibly different delay requirements sharing a wireless channel resource, we derive the resource allocation and rate adaptation policy that maximizes the sum video quality. Resource allocation adapts the partitioning of the wireless channel resources across the users and rate adaptation adapts the video source rate of each user. We show that the optimal operating point per user is such that the rate-distortion slope is the inverse of the supported video source rate per unit bandwidth. The maximum source rate per unit bandwidth is a fundamental measure of the number of video bits per channel use that can be delivered subject to the QoS requirement and we refer to it as the \emph{source spectral efficiency}. Next, we solve the alternative problem of fairness-based resource allocation whereby the objective is to maximize the minimum video quality across users and contrast the solution with the sum quality maximizing policy.

\subsubsection{Maximal user subset scheduling}

We derive a scheduling policy to select a subset of users such that all scheduled users can meet their statistical delay requirement. We show that the optimal scheduling policy can be obtained in polynomial time in the number of users and it involves computing the minimum resource allocation required by each user to support their QoS requirement, using it as a sorting criterion, and scheduling the first sorted users such that the sum of their minimum resource requirement does not exceed the total available resources. Under the fairness constraint, a similar solution is obtained with the major difference that the sorting criterion is the video quality corresponding to the minimum rate representation of the video sequence.

\subsubsection{Statistical QoS-based user admission}

We extend the problem to accommodate dynamically changing networks whereby users can request new sessions with certain target QoS requirement. We derive the admission policy that ensures the admitted user can meet the statistical delay constraint without jeopardizing any of the other users' QoS requirements. The user admission policy is derived under both sum quality-maximizing resource allocation as well as fairness-based resource allocation and it depends on the minimum rate representation available for the video sequences corresponding to each user as well as the QoS requirements of each user.

\begin{figure*}[t!]
    \centering
    \includegraphics[width=15cm]{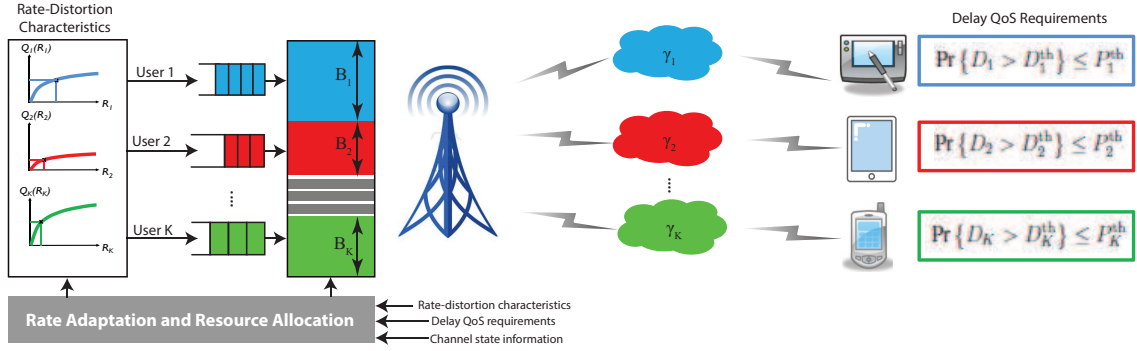}
    \caption{\label{fig:block_diagram} Proposed system block diagram for quality-driven resource allocation and rate adaptation of delay-constrained video streams.}
\end{figure*}

\subsection{Related Work}

The effective capacity link layer model characterizes the capacity of the wireless channels in the presence of queues using QoS exponents that describe the decay rate of the queue length tail probability and characterize a corresponding statistical delay bound. For scalable video transmission, effective capacity analysis is applied in \cite{stat_qos_uni_multi} to provide statistical delay bounds for scalable video transmission over unicast and multicast links. Considering energy-efficiency as a target objective in an ad-hoc network with possible multihop transmissions, \cite{Khalek_TMC} derives
energy-efficient transmission schemes such that the end-to-end delay bounds are satisfied. Power and rate adaptation with effective capacity-driven quality of service provisions is considered in \cite{tang_qos}. While addressing statistical delay bounds for video transmission, no previous work directly optimizes video quality, but instead minimizes resource utilization \cite{stat_qos_uni_multi,tang2008cross}, maximizes rate or throughput \cite{tang_qos,vassaki2012effective,ma2012power}, or minimizes energy consumption \cite{Khalek_TMC,tang2008cross}.

Multiuser scheduling and resource allocation for video transmission has been studied in \cite{joseph_variability,pahalawatta2007content,dawei_crosslayerOFDMA,maani2008resource,ji2009scheduling}. In \cite{joseph_variability}, the knowledge of the variations in the dependence of perceived video quality to the compression rate is utilized for resource allocation to improve video QoE across multiple users. Accounting for users' QoE variability in this manner allows adapting resource allocation in a content-aware manner. An asymptotically optimal online algorithm for optimizing users' QoE is proposed that allows realizing tradeoffs across mean QoE, variance in QoE, and fairness. This work applies for stored video streaming as it assumes reliable transport and large buffer-ahead which enables tolerating large delays. In \cite{pahalawatta2007content}, a cross-layer packet scheduling scheme that streams pre-encoded video over wireless downlink packet access networks to multiple users is presented. A scheduling scheme is used in which data rates are dynamically adjusted based on channel quality as well as the gradients of a video quality utility function. In addition to video distortion, the design of the utility function takes into account decoder error concealment. In \cite{dawei_crosslayerOFDMA}, the problem of multiuser resource allocation for uplink OFDMA is studied taking into account the rate-distortion characteristics of different users. A subcarrier assignment and power allocation algorithm is devised to minimize the average video distortion among users in the system based on the CSI and rate-distortion. In \cite{maani2008resource}, a cross-layer resource allocation and packet scheduling scheme is developed taking into account the time-varying nature of the wireless channels to minimize the expected distortion of the received sequence. Complex error concealment is taken into account in estimating video distortion and the gradients of the distortion are used to efficiently allocate resources across users. In \cite{ji2009scheduling}, the problem of scheduling and resource allocation for multiuser video streaming over downlink OFDM channels is addressed for SVC with quality and temporal scalability. A scheduling algorithm prioritizes the transmissions of different users by considering video contents, deadline requirements, and transmission history.


A significant body of literature is devoted to point-to-point optimizations applied to real-time video transmission. Such adaptive video transmission techniques can be roughly categorized into joint source-channel coding (JSCC) \cite{Girod_JSCC,Kondi_JSCC,zhang2007joint,Khalek_Globecom}, unequal error protection (UEP) \cite{Kim_UEP,Gallant_UEP,Khalek_JSAC,Khalek_MLSP,Khalek_Globecom}, prioritized scheduling \cite{Schaar_scheduling}, and loss visibility-based prioritization \cite{LinPrioritization,Khalek_EUSIPCO,khalek2012MM}. While these point-to-point optimizations are useful for improving video quality and error resilience for individual video streams, they don't capture hybrid user dynamics in terms of rate-distortion behavior and differentiated QoS requirements which is the key focus of this paper.

Previous work on video transmission in a multiuser setup focuses on stored video use cases (e.g. \cite{joseph_variability,pahalawatta2007content,dawei_crosslayerOFDMA,maani2008resource}) in which content can be buffered ahead opportunistically and large delays can be tolerated. In that scenario, the problem reduces to rate-distortion optimizations and buffer management policies. For real-time video, however, such as live streaming or video conferencing, whereby the content in not pre-encoded, large buffers are not feasible and stringent delay constraints need to be guaranteed. For that important use case, no previous work addresses the problem of optimizing a video quality-based utility function while guaranteeing a statistical delay bound per user.


\subsection{Paper Organization}

The rest of the paper is organized as follows: We present the system model in Section \ref{sys_model}. The background for providing statistical delay bounds using the theory of effective capacity is presented in Section \ref{delayQoS}. In Section \ref{sec:resource_alloc}, we solve the problem of quality-maximizing resource allocation under statistical delay constraints as well as fairness-based resource allocation. In Section \ref{sec:scheduling}, we present the maximal subset scheduling solution and the user admission criterion. In Section \ref{sec:Alg}, we present the algorithm for joint scheduling and resource allocation under the quality-maximizing policy and the fairness-based policy. We present results and analysis of the corresponding gains in Section \ref{sec:results}. Finally, concluding remarks are provided in Section \ref{sec:conc}. 

\section{System Model}\label{sys_model}

Consider the downlink of a base station where $K$ video users orthogonally share a bandwidth $B$ Hz. Each user $k$ is allocated a bandwidth $B_k$ Hz such that $\sum_{k=1}^{K}{B_k} = B$. The individual user bandwidths are assumed narrowband so that the corresponding wireless channel experiences flat fading. We note that in a wideband system, with orthogonal frequency division multiple access (OFDMA), this assumption can be ensured by allocating adjacent subcarriers to individual users.

The channel coherence time is $T$. Further, the timescale of video rate adaptation is much larger than $T$. This is typically the case in practice because the rate of the video source is adapted at the GoP timescale which is typically on order of a second. The channel variation is assumed to be in the milliseconds timescale. For slow fading channels whereby the channel varies at the same timescale as the source, statistical delay guarantees are not needed since the service rate is almost deterministic. Thus, the challenging case of interest is when the channel state variation is considerably faster the source rate variation.

Since resource allocation and rate adaptation are done at the source variation timescale and the video source experiences the ergodic capacity of the channel, the base station (BS) or access point (AP) only require channel distribution information (CDI). Let $f_k(\gamma)$ represent the SNR distribution for user $k$. Moreover, $\{\gamma_k\}$ is modeled as an ergodic and stationary block-fading process uncorrelated among consecutive packets $j$. We assume that each user experiences Gaussian noise such that, with capacity-achieving codes, the instantaneous transmission rate for user $k$ is $C_{k} = B_k \log(1+\gamma_k)$.




The video segment intended for user $k$ is transmitted at rate $R_k$. A queue is inserted for each user stream to absorb the mismatch between the arrival and service rate due to the channel variations. The video stream for each user $k$ is characterized by a rate distortion model $Q_k(R_k)$ that determines the mapping from the source rate to the video perceptual quality. The rate-quality mapping function $Q_k(.)$ is concave and continuous. In practice, rate distortion curves corresponding to practical video codecs always follow this concave behavior because diminishing returns in quality are achieved as the rate increases.

The corresponding system block diagram is shown in Figure \ref{fig:block_diagram}. The three major per-user dynamics are: (1) rate-distortion characteristics $Q_k(R_k)$, (2) channel statistics $f_k(\gamma)$, and (3) delay QoS requirements $\{D_{k,\mathrm{th}},P_{k,\mathrm{th}}\}$. We focus on resource allocation and rate adaptation for users served by a single cell. We note, however, that the results in the paper apply to a system with other-cell interference if the interference is treated as noise where $f_k(\gamma)$ then represents the distribution of the signal to noise and interference ratio (SINR).



\section{Statistical Delay Bounds}\label{delayQoS}

In this section, we describe the procedure for providing delay guarantees by characterizing link-level QoS metrics according to the effective capacity link layer model.

\subsection{Queuing Model for Video Transmission}


A separate queue is maintained for each video stream at the base station. Given the SNR distribution $f_k(\gamma)$ for each user, the objective is to adapt the source rates $R_k$ and the bandwidth allocation $B_k$ such that the following QoS constraint is satisfied

\begin{equation}\label{eqn:delay-bound-simplified}
\textrm{Pr}\left\{{D_{k}}>D_{k}^\mathrm{th}\right\} \le P_{k}^{\mathrm{th}}~~~\forall~k
\end{equation}

\noindent where $D_{k}$ is the queuing and transmission delay for user $k$ video stream, $D_{k}^{\mathrm{th}}$ is the statistical delay-bound, and $P_{k}^{\mathrm{th}}$ is the target delay-bound violation probability.


The behavior of the queue-length process in queuing-based communication networks is extensively treated in
\cite{queue_length_delay}. For an ergodic and stationary arrival and service processes, the queue length at time $t$ of each queue can be bounded exponentially as $t\rightarrow \infty$

\vspace{-0.3cm}
\begin{equation}\label{eqn:queue-length}
\textrm{Pr}\left\{l_k(t)>l_{k}^{\mathrm{th}}\right\}\doteq e^{-\theta_k l_{k}^{\mathrm{th}}} \ \ \ \ \ \ 1\le k\le K
\end{equation}

\noindent where $l_{k}$ is the queue length at queue $k$ and $l_{k}^{\mathrm{th}}$ is the queue-length threshold. The
parameter $\theta_k$, termed the \emph{QoS exponent}, determines the decay rate and is used to characterize delay. More stringent QoS requirements are characterized by larger $\theta_k$ while looser QoS requirements require smaller $\theta_k$.

\subsection{Effective Capacity for Statistical Delay Bounds}

The effective capacity (EC) channel model  captures a generalized link-level
capacity notion of the fading channel by characterizing wireless channels in terms of functions that can be easily mapped to link-level QoS metrics, such as delay-bound violation probability. Thus, it is a convenient tool for designing QoS provisioning mechanisms \cite{eff_cap}, \cite{stat_qos_uni_multi}.

We denote by $\mathcal{C}_{k}(\theta_k)$ and $\mathcal{A}_{k}(\theta_k)$ the effective capacity and effective bandwidth functions, respectively, for the $k^{th}$ user. Given an arrival process $\{A_{k}\}$, its effective bandwidth, denoted by $\mathcal{A}_{k}(\theta_k)$ (bits/GoP), is defined as the minimum constant service rate required to guarantee a
specified QoS exponent $\theta_k$. In contrast, for a given service process $\{C_{k}\}$, its effective capacity, denoted by $\mathcal{C}_{k}(\theta_k)$ (bits/GoP), is defined as the maximum
constant arrival rate which can be supported by $\{C_{k}\}$ subject to the specified QoS exponent $\theta_k$.

The effective capacity can be thought of as the capacity of the underlying channel from the perspective of upper protocol layers. The effective capacity theory states that considering the delay as the performance metric instead of the queue length in \eqref{eqn:queue-length}, the QoS guarantee can be written equivalently as a function of the effective capacity as follows

\vspace{-0.3cm}
\begin{equation}\label{eqn:qos-guarantee}
\textrm{Pr}\left\{D_k(t)>D_{k}^{\mathrm{th}}\right\}\doteq e^{-\theta_k \mathcal{C}_{k}(\theta_k) D_{k}^{\mathrm{th}}} \ \ \ \ \ \ 1\le k\le K.
\end{equation}

\noindent Moreover, for a stationary and ergodic service process $\{C_{k}\}$ that is uncorrelated across time frames, the effective capacity can be expressed as \cite{tang_qos}

\vspace{-0.3cm}
\begin{eqnarray}
\mathcal{C}_{k}(\theta_k) &=& -\frac{1}{\theta_k}\ln\left(\mathbb{E}_{\boldsymbol{\gamma}}\{e^{-\theta_k C_{k}}\}\right)\label{eqn:effcap1}\nonumber
\\&=& -\frac{1}{\theta_k}\ln\left(\mathbb{E}_{\boldsymbol{\gamma}}\{e^{-\theta_k B_k T \log(1+\gamma_k)}\}\right)\label{eqn:effcap2}\cdot
\end{eqnarray}

\noindent To provide the QoS guarantee $\theta_k$ for user $k$, the effective capacity on the $k^{th}$ link should be equal to the effective bandwidth \cite{eff_band,stat_qos_uni_multi,stat_qos-cross_layer}, i.e.,

\vspace{-0.3cm}
\begin{equation}\label{eqn:effband-cap}
\mathcal{C}_{k}(\theta_k) = \mathcal{A}_{k}(\theta_k)~~~~~\forall k=1,\cdots,K.\nonumber
\end{equation}

\noindent The effective bandwidth for the arrival process described above is $\mathcal{A}_{k}(\theta_k) = R_k T$. Thus, for the QoS constraint in \eqref{eqn:delay-bound-simplified} to be satisfied, we need to ensure that $\mathcal{C}_{k}(\theta_k)\ge \bar{\mathcal{C}}_{k} = R_k T$. The QoS constraint reduces to

\begin{equation}\label{eqn:max_source_rate}
R_k  \le -\frac{1}{T\theta_k}\ln\left(\mathbb{E}_{\boldsymbol{\gamma}}\{e^{-\theta_k B_k T \log(1+\gamma_k)}\}\right).\nonumber
\end{equation}

\noindent Furthermore, we can find $\theta_k$ by solving $\mathcal{C}_{k}(\theta_k) = R_k T$ using \eqref{eqn:qos-guarantee} as follows

\vspace{-0.2cm}
\begin{equation}\label{eqn:theta}
\theta_k = \frac{\ln(1/P_{k}^{\mathrm{th}})}{T R_k D_{k}^{\mathrm{th}}}.
\end{equation}

\section{Quality-driven Resource Allocation and Rate Adaptation}\label{sec:resource_alloc}

In this section, we formulate and solve the problem of resource allocation and rate adaptation for maximizing the sum video quality across users subject to the statistical delay constraint.

\vspace{-0.3cm}

\subsection{Sum Quality-Maximizing Policy}\label{formulate-general-allocation}

The problem of optimizing the source rate vector $\mathbf{R} = \{R_k\}_{k=1}^{K}$ and the bandwidth allocation vector $\mathbf{B} = \{B_k\}_{k=1}^{K}$ to maximize the sum video quality subject to a statistical delay constraint per user is formulated as follows

\vspace{-0.3cm}

\begin{eqnarray}
&\hspace{-0.3cm}\max_{\mathbf{R},\mathbf{B}} & \sum_{k=1}^{K}{Q_k(R_k)}\label{eqn:obj}\nonumber
\\&\hspace{-0.3cm}\mathrm{s.t.} & \sum_{k=1}^K{B_k} = B;~B_k \ge 0~~\forall~k\label{eqn:constr1}
\\&\hspace{-0.3cm}& \textrm{Pr}\left\{{D_{k}}>D_{k}^{\mathrm{th}}\right\} \le P_{k}^{\mathrm{th}}~~~\forall~k\label{eqn:constr2}
\end{eqnarray}

\noindent Using the theory of effective capacity, Lemma 1 provides the maximal video source rate supported by each user subject to the statistical delay constraint.

\begin{table}[b]\scriptsize\label{tbl:notation}\centering
  \caption{Commonly used notation}\vspace{-0.3cm}
\begin{tabular}{|c|l|c|l|}
  \hline
  \multicolumn{2}{|c|}{Video source / QoS metrics (User $k$)} & \multicolumn{2}{|c|}{Wireless Channel / System} \\   \hline
  $D_{k}^{\mathrm{th}}$ & Delay constraint & $K$ & Number of video users \\
  $P_{k}^{\mathrm{th}}$ & Delay violation probability & $B_k$ & User $k$ BW allocation\\
  $\theta_k$ & QoS exponent &$B$ & Total bandwidth\\
  $R_k$ & Source rate&   $\mathcal{C}_k(\theta_k)$ & User $k$ effective capacity \\
  $R_k^{\min}$ & Minimum rate representation & $f_k(\gamma)$ & Fading distribution \\
$Q_k(R_k)$ & Rate-quality mapping & $T$ & Channel coherence time\\
   $d_k$ & BW-QoS exponent product&  $R_k/B_k$ & Source spectral efficiency \\
  \hline
\end{tabular}
\end{table}

\begin{Lemma1}
For a given bandwidth allocation vector $\mathbf{B}$ and set of QoS exponents $\boldsymbol{\theta}$, the optimal source rate $R_k$ for user $k$ is

\begin{eqnarray}
R_k^*(B_k,\theta_k) &=& -\frac{1}{T\theta_k}\ln\left(\mathbb{E}_{\boldsymbol{\gamma}}\left\{e^{-\theta_k B_k T \log(1+\gamma_k)}\right\}\right)\nonumber
\\&=& -\frac{1}{T\theta_k}\ln\left(\mathbb{E}_{\boldsymbol{\gamma}}\left\{(1+\gamma_k)^{-\frac{\theta_k B_k T}{\ln(2)}}\right\}\right).\nonumber
\end{eqnarray}

\end{Lemma1}

\begin{proof} Since $Q_k(R_k)$ is increasing, increasing the source rate for any user improves the objective function. Furthermore, the delay bound violation probability $\textrm{Pr}\left\{{D_{k}}>D_{k}^{\mathrm{th}}\right\}$ is increasing in the source rate. Thus, the optimal solution must satisfy  $\textrm{Pr}\left\{{D_{k}}>D_{k}^{\mathrm{th}}\right\} = P_{k}^{\mathrm{th}}~\forall k$. The equality in the statistical delay bound implies that the effective capacity constraint $\mathcal{C}_{k}(\theta_k)\ge \bar{\mathcal{C}}_{k} = R_k T$ is satisfied with equality, that is, $\mathcal{C}_{k}(\theta_k) = R_k T$. Plugging in $\mathcal{C}_{k}(\theta_k)$ from \eqref{eqn:effcap2}, we obtain $R_k^*(B_k,\theta_k) = -\ln\left(\mathbb{E}_{\boldsymbol{\gamma}}\{e^{-\theta_k B_k T \log(1+\gamma_k)}\}/{T\theta_k}\right)$ and the result follows.
\end{proof}

Since the QoS exponent in \eqref{eqn:theta} is also a function of the source rate which in turn is a function of the bandwidth allocation, it cannot be computed beforehand. Thus, we compute $\theta_k$ jointly with $B_k$ as part of the optimization problem. The following Lemma re-writes the delay constraint in terms of the QoS exponent and the bandwidth allocation using the effective capacity expression for an uncorrelated channel.

\begin{Lemma2}
The statistical delay constraint \eqref{eqn:constr2} can be written equivalently as

\vspace{-0.3cm}

\begin{equation}
\mathbb{E}_{\boldsymbol{\gamma}}\left\{(1+\gamma_k)^{-\frac{\theta_k B_k T}{\ln(2)}}\right\} = p_k\nonumber
\end{equation}

\noindent where $p_k = {P_{k}^{\mathrm{th}}}^{1/D_{k}^{\mathrm{th}}}$. Furthermore, the solution to the problem requires that $\theta_k^* = d_k/B_k^*$ where the \emph{bandwidth-QoS exponent product} $d_k$ is a constant dependent only on the channel distribution information and the delay constraint.
\end{Lemma2}

\begin{proof}
From Lemma 1, the statistical delay constraint should be satisfied with equality. Thus, we can rewrite the statistical delay constraint as follows

\vspace{-0.3cm}

\begin{eqnarray}
e^{-\theta_k \mathcal{C}_{k}(\theta_k)D_{k}^{\mathrm{th}}} &=& P_{k}^{\mathrm{th}}\nonumber\\
\theta_k T R_k^*(B_k,\theta_k) D_{k}^{\mathrm{th}} &=& -\ln(P_{k}^{\mathrm{th}})\label{lemma2-step1}\nonumber\\
\ln\left(\mathbb{E}_{\boldsymbol{\gamma}}\left\{(1+\gamma_k)^{-\frac{\theta_k B_k T}{\ln(2)}}\right\}\right)  &=& \frac{\ln(P_{k}^{\mathrm{th}})}{D_{k}^{\mathrm{th}}}\label{lemma2-step2}
\\ \mathbb{E}_{\boldsymbol{\gamma}}\left\{(1+\gamma_k)^{-\frac{\theta_k B_k T}{\ln(2)}}\right\}  &=& {P_{k}^{\mathrm{th}}}^{1/D_{k}^{\mathrm{th}}} = p_k\label{lemma2-step3}
\end{eqnarray}

\noindent where \eqref{lemma2-step2} follows using Lemma 1. Furthermore, since $p_k$ in \eqref{lemma2-step3} is a constant independent of the channel statistics and the bandwidth allocation, it follows that $\theta_k B_k$ should be constant. Thus, $\theta_k^* = d_k/B_k^*$ for some constant $d_k$.
\end{proof}

Applying Lemma 1 and Lemma 2, we rewrite the resource allocation problem equivalently as follows

\vspace{-0.3cm}

\begin{eqnarray}
&\hspace{-0.3cm}\min_{\mathbf{B},\mathbf{\theta}} & -\sum_{k=1}^{K}{Q_k\left(-\frac{1}{T\theta_k}\ln\left(\mathbb{E}_{\boldsymbol{\gamma}}\{(1+\gamma_k)^{-\frac{\theta_k B_k T}{\ln(2)}}\}\right)\right)}\label{eqn:obj1}\nonumber
\\&\hspace{-0.3cm}\mathrm{s.t.} & \sum_{k=1}^K{B_k} = B\label{eqn:constr1-1}
\\&\hspace{-0.3cm}& \mathbb{E}_{\boldsymbol{\gamma}}\left\{(1+\gamma_k)^{-\frac{\theta_k B_k T}{\ln(2)}}\right\} = p_k~~~\forall~k\label{eqn:constr2-1}
\\&\hspace{-0.3cm}& B_k \ge 0~~\forall~k.\label{eqn:constr3-1}
\end{eqnarray}

\noindent First, we prove the convexity of the problem above both in $\mathbf{B}$ and $\mathbf{\theta}$ in Lemma 3.

\begin{Lemma3}
The problem is convex in $\mathbf{B}$ and $\boldsymbol{\theta}$. Furthermore, a feasible point always exists if $Q_k(R_k)$ is defined for every $R_k\ge 0$.
\end{Lemma3}

\begin{proof}
First, we show convexity of the objective function in $B_k$. Since $-Q_k(R_k)$ is convex in $R_k$ and non-increasing, it suffices to show that $R_k = -\frac{1}{T\theta_k}\ln\left(\mathbb{E}_{\boldsymbol{\gamma}}\{(1+\gamma_k)^{-\frac{\theta_k B_k T}{\ln(2)}}\}\right)$ is concave in $B_k$. We have


\begin{eqnarray}
\alpha R_k(B_k^1) \hspace{-0.3cm}&+& \hspace{-0.3cm}(1-\alpha) R_k(B_k^2)\nonumber
\\  &&\hspace{-2.1cm} = \frac{-1}{T\theta_k}\left[\alpha\ln\hspace{-0.1cm}\left(\mathbb{E}_{\boldsymbol{\gamma}}\hspace{-0.1cm}\left\{(1+\gamma_k)^{-\frac{\theta_k B_k^1 T}{\ln(2)}}\right\}\right)\right.\nonumber
\\  &&\hspace{1cm} +\left. (1-\alpha)\ln\hspace{-0.1cm}\left(\mathbb{E}_{\boldsymbol{\gamma}}\hspace{-0.1cm}\left\{(1+\gamma_k)^{-\frac{\theta_k B_k^2 T}{\ln(2)}}\right\}\right)\right]\nonumber
\\ &&\hspace{-2.1cm} = \frac{-1}{T\theta_k}\ln\hspace{-0.1cm}\left(\hspace{-0.1cm} \mathbb{E}_{\boldsymbol{\gamma}}\hspace{-0.1cm}\left\{(1+\gamma_k)^{-\frac{\theta_k B_k^1 T}{\ln(2)}}\right\}^{\hspace{-0.05cm}\alpha} \hspace{-0.1cm} \mathbb{E}_{\boldsymbol{\gamma}}\hspace{-0.1cm}\left\{(1+\gamma_k)^{-\frac{\theta_k B_k^2 T}{\ln(2)}}\right\}^{\hspace{-0.05cm} 1-\alpha}\hspace{-0.05cm} \right).\nonumber
\end{eqnarray}

\noindent Now, apply Lyaponuv's inequality which states that $ \mathbb{E}[|Y|^s]^r \times \mathbb{E}[|Y|^l]^{1-r} \ge \mathbb{E}[|Y|^m]$ where $r = (l-m)/(l-s)$ and $0<s<m<l$. We assume $B_k^1 < B_k^2$ w.l.o.g. and apply the inequality for $|Y| = (1+\gamma_k)^{-1}$, $r = \alpha$, $s = \theta_k T B_k^1/\ln(2)$, and $l = \theta_k T B_k^2/\ln(2)$. Thus, $\alpha = (\theta_k T B_k^2/\ln(2) - m) / (\theta_k T B_k^2/\ln(2) - \theta_k T B_k^1/\ln(2))$. Solving for $m$, we obtain $m = \frac{\theta_k T }{\ln(2)}(\alpha B_k^1 + (1-\alpha) B_k^2)$. Applying the inequality, we obtain


\begin{eqnarray}
\alpha R_k(B_k^1) &+& (1-\alpha) R_k(B_k^2)\nonumber
\\ && \hspace{-1.3cm} \le- \frac{1}{T\theta_k}\left[\ln\left(\mathbb{E}_{\boldsymbol{\gamma}}\left\{(1+\gamma_k)^{-\frac{\theta_k T }{\ln(2)}(\alpha B_k^1 + (1-\alpha) B_k^2)}\right\} \right)\right]\nonumber
\\ && \hspace{-1.3cm}=    R_k(\alpha B_k^1 + (1-\alpha) B_k^2).\nonumber
\end{eqnarray}

\noindent Thus, $R_k(B_k)$ is concave in $B_k$ and the objective function is convex in $B_k$. Next, we show that the objective function is also convex in $\theta_k$, It is clear that the component $\ln\left(\mathbb{E}_{\boldsymbol{\gamma}}\{(1+\gamma_k)^{-\frac{\theta_k B_k T}{\ln(2)}}\}\right)$ is convex in $\theta_k$ as well by the exact arguments used above. Thus, we have $R_k(\theta_k) = -\frac{1}{T\theta_k} f(\theta_k)$ where $f(\theta_k)$ is convex and non-increasing in $\theta_k$. Also, $\frac{1}{T\theta_k}$ is convex and non-increasing. The product of two convex and non-increasing functions is convex. Thus, $\frac{1}{T\theta_k} f(\theta_k)$ is convex and $R_k(\theta_k)$ is concave in $\theta_k$. Applying the same arguments, we can further show that the constraint \eqref{eqn:constr2-1} is convex in $B_k$ and $\theta_k$. Thus, the optimization problem is convex and has a unique solution if it feasible.

We further show that the problem is always feasible if $Q_k(R_k)$ is defined for every $R_k \ge 0$, i.e., the video source rate can be arbitrarily reduced. Using Lemma 2, we can always find a positive $d_k = B_k \theta_k$ that satisfies the statistical delay constraint. For any $B_k$, $\theta_k$ that satisfies $B_k \theta_k = d_k$, we use Lemma 1 to find a corresponding source rate $R_k$ that can be supported subject to the delay bound. To summarize, given any delay bound and channel conditions, there exists a small enough source rate such that the delay bound is met.
\end{proof}

Theorem 1 provides the optimal resource allocation solution by showing that the optimal operating point per user is such that the rate-distortion slope is the inverse of the supported video source rate per unit bandwidth. The maximum source rate per unit bandwidth is a fundamental measure of the number of video bits per channel use that can be delivered subject to the QoS requirement which we refer to as the \emph{source spectral efficiency}.

\begin{Thm1}
\textbf{Sum Quality-Maximizing Resource Allocation:} The optimal bandwidth allocation $B_k$ and source rate $R_k$ for each user $k$ is such that

\begin{equation}
\underbrace{\frac{\partial Q_k(R_k^*(B_k^*,\theta_k^*))}{\partial R_k^*(B_k^*,\theta_k^*)}}_{\textrm{Rate~distortion~slope}}\times \underbrace{\frac{R_k^*(B_k^*,\theta_k^*)}{B_k^*}}_{\textrm{Source~spectral~efficiency}}  = \rho\nonumber
\end{equation}

\noindent where $R_k^*(B_k,\theta_k) = -\frac{1}{T\theta_k}\ln\left(\mathbb{E}_{\boldsymbol{\gamma}}\{(1+\gamma_k)^{-\frac{\theta_k B_k T}{\ln(2)}}\}\right)$ and $\rho$ is a constant chosen such that $\sum_{k=1}^{K}{B_k} = B$. Further, $R_k^*(B_k,\theta_k^*)/B_k$ is independent of $B_k$ and is only a function of the fading distribution $f_k(\gamma)$, the delay bound $D_{k}^\mathrm{th}$, and the delay bound violation probability $P_{k}^\mathrm{th}$.
\end{Thm1}

\begin{proof}
We write the Lagrangian function for the resource allocation problem and the Karush-Kuhn-Tucker (KKT) conditions to derive the optimal solution. The Lagrangian function $L(\mathbf{B},\boldsymbol{\theta},\rho,\boldsymbol{\lambda},\boldsymbol{\mu})$ of the resource allocation problem is as follows

\vspace{-0.5cm}

\small
\begin{eqnarray}
&&\hspace{-0.7cm}L(\mathbf{B},\boldsymbol{\theta},\rho,\boldsymbol{\lambda},\boldsymbol{\mu}) = -\sum_{k=1}^{K}{Q_k\left(\frac{-1}{T\theta_k}\ln\left(\mathbb{E}_{\boldsymbol{\gamma}}\{(1+\gamma_k)^{-\frac{\theta_k B_k T}{\ln(2)}}\}\right)\right)} \nonumber \\ &&\hspace{-0.7cm}- \rho\hspace{-0.1cm}\left(\sum_{k=1}^K{B_k} - B\right)\hspace{-0.1cm}+\hspace{-0.1cm} \sum_{k=1}^{K}{\lambda_k \hspace{-0.1cm}\left(\mathbb{E}_{\boldsymbol{\gamma}}\hspace{-0.1cm}\left\{(1+\gamma_k)^{-\frac{\theta_k B_k T}{\ln(2)}}\right\} \hspace{-0.1cm}- p_k\hspace{-0.1cm}\right)} \hspace{-0.1cm}-\hspace{-0.1cm} \sum_{k=1}^{K}{\mu_k B_k}\nonumber
\end{eqnarray}
\normalsize

\noindent where $\rho$, $\boldsymbol{\lambda} = \{\lambda_k\}_{k=1}^K$, and $\boldsymbol{\mu} = \{\mu_k\}_{k=1}^K$ are the Lagrange multipliers corresponding to constraints \eqref{eqn:constr1-1}, \eqref{eqn:constr2-1}, and \eqref{eqn:constr3-1} respectively. The KKT conditions are

\vspace{-0.6cm}

\begin{eqnarray}\hspace{-1cm}
\begin{cases}\frac{\partial L}{\partial B_k }\big|_{B_k=B_k^*}=0, \frac{\partial L}{\partial \theta_k }\big|_{\theta_k=\theta_k^*}=0~~\forall k\label{eqn:KKT1}
\\\mu_k^*\ge 0, B_k^* \ge 0, \mu_k^* B_k^* = 0~~\forall k\label{eqn:KKT2}
\\ \sum_{k=1}^{K}{B_k^*} = B \label{eqn:KKT3}
\\ \mathbb{E}_{\boldsymbol{\gamma}}\left\{(1+\gamma_k)^{-\frac{\theta_k^* B_k^* T}{\ln(2)}}\right\} = p_k~~~\forall~k.\label{eqn:KKT4}
\end{cases}\nonumber
\end{eqnarray}

\noindent Taking the derivative of $L(\mathbf{B},\boldsymbol{\theta},\rho,\boldsymbol{\lambda},\boldsymbol{\mu})$ with respect to $B_k$, we obtain

\vspace{-0.6cm}

\small
\begin{eqnarray}
\frac{\partial L}{\partial B_k } =&& \hspace{-0.7cm}\frac{\partial Q_k(R_k^*(B_k,\theta_k))}{\partial R_k^*(B_k,\theta_k)}\times  \frac{\mathbb{E}_{\boldsymbol{\gamma}}\left\{(1+\gamma_k)^{-\frac{\theta_k B_k T}{\ln(2)}}\log(1+\gamma_k)\right\}}{\mathbb{E}_{\boldsymbol{\gamma}}\left\{(1+\gamma_k)^{-\frac{\theta_k B_k T}{\ln(2)}}\right\}}\nonumber
\\ &&\hspace{-0.8cm}-\rho - \lambda_k \theta_k T \left(\mathbb{E}_{\boldsymbol{\gamma}}\left\{(1+\gamma_k)^{-\frac{\theta_k B_k T}{\ln(2)}}\log(1+\gamma_k)\right\}\right)- \mu_k\nonumber
\\&&\hspace{-1.7cm}= \theta_k \underbrace{T \mathbb{E}_{\boldsymbol{\gamma}}\left\{(1+\gamma_k)^{\frac{-\theta_k B_k T}{\ln(2)}}\hspace{-0.05cm}\log(1+\gamma_k)\right\}\hspace{-0.1cm}\left(\frac{\partial Q_k(R_k^*)/\partial R_k^*}{\theta_k T p_k} - \lambda_k\right)}_{g(B_k,\theta_k,\lambda_k)} \nonumber\\&&\hspace{-0.8cm}-\rho - \mu_k\nonumber
\end{eqnarray}


\normalsize

\noindent where we used the last KKT condition and we defined $g(B_k,\theta_k,\lambda_k)$ as stated above. Thus, we have

\begin{equation}
\frac{\partial L}{\partial B_k } =\theta_k {g(B_k,\theta_k,\lambda_k)} -\rho - \mu_k.\nonumber
\end{equation}

\noindent Note that $\partial Q_k(R_k^*)/\partial R_k^*$ is the slope of the rate distortion function at $R_k^*$. Now, we take the derivative of the Lagrangian with respect to $\theta_k$ as follows

\small
\begin{eqnarray}
\frac{\partial L}{\partial \theta_k } =&& \hspace{-1cm}\frac{\partial Q_k(R_k^*)}{\partial R_k^*}\left[ \frac{\ln\left(p_k\right)}{T\theta_k^2} + \frac{B_k}{\theta_k p_k}\mathbb{E}_{\boldsymbol{\gamma}}\left\{(1+\gamma_k)^{-\frac{\theta_k B_k T}{\ln(2)}}\log(1+\gamma_k)\right\}\right]\nonumber
\\ &-& \lambda_k B_k T \left(\mathbb{E}_{\boldsymbol{\gamma}}\left\{(1+\gamma_k)^{-\frac{\theta_k B_k T}{\ln(2)}}\log(1+\gamma_k)\right\}\right)\nonumber
\\&&\hspace{-1.8cm} =B_k \underbrace{T \mathbb{E}_{\boldsymbol{\gamma}}\left\{(1+\gamma_k)^{-\frac{\theta_k B_k T}{\ln(2)}}\log(1+\gamma_k)\right\}\left(\frac{\partial Q_k(R_k^*)/\partial R_k^*}{\theta_k T p_k} - \lambda_k \right)}_{g(B_k,\theta_k,\lambda_k)} \nonumber\\ &+& \frac{\partial Q_k(R_k^*)}{\partial R_k^*} \frac{\ln\left(p_k\right)}{T\theta_k^2} \nonumber
\\&& \hspace{-1.8cm}=B_k {g(B_k,\theta_k,\lambda_k)} + \frac{\partial Q_k(R_k^*)}{\partial R_k^*} \frac{\ln\left(p_k\right)}{T\theta_k^2}.\nonumber
\end{eqnarray}
\normalsize

\noindent Substituting $\frac{\partial L}{\partial B_k }\big|_{B_k=B_k^*}=0$ and $\frac{\partial L}{\partial \theta_k }\big|_{\theta_k=\theta_k^*}=0$, we obtain the following set of equations

\begin{equation}
\begin{cases}
\theta_k^* {g(B_k^*,\theta_k^*,\lambda_k^*)} -\rho^* - \mu_k^* = 0\\
B_k^* {g(B_k^*,\theta_k^*,\lambda_k^*)} + \frac{\partial Q_k(R_k^*(B_k^*,\theta_k^*))}{\partial R_k^*(B_k^*,\theta_k^*)} \frac{\ln\left(p_k\right)}{T\theta_k^{*2}} = 0.
\end{cases}\nonumber
\end{equation}

\noindent Substituting $g(B_k^*,\theta_k^*,\lambda_k^*) = (\mu_k^* + \rho^*)/\theta_k^*$ into the second equation, we obtain

\begin{equation}
\frac{B_k^* (\mu_k^* + \rho^*)}{\theta_k^*} + \frac{\partial Q_k(R_k^*(B_k^*,\theta_k^*))}{\partial R_k^*(B_k^*,\theta_k^*)} \frac{1}{T\theta_k^{*2}}\ln\left(p_k\right) = 0.\nonumber
\end{equation}

\noindent Equivalently,

\begin{equation}
\frac{\partial Q_k(R_k^*(B_k^*,\theta_k^*))}{\partial R_k^*(B_k^*,\theta_k^*)} \ln\left(p_k\right) = -T B_k^* \theta_k^* (\rho^* + \mu_k^*).\nonumber
\end{equation}

\noindent To obtain another form of the expression that has a useful interpretation, we substitute  $R_k^*(B_k,\theta_k) =-\ln\left(\mathbb{E}_{\boldsymbol{\gamma}}\left\{(1+\gamma_k)^{-\frac{\theta_k^* B_k^* T}{\ln(2)}}\right\}\right)/(T\theta_k)$. Thus, we have

\begin{equation}
\frac{\partial Q_k(R_k^*(B_k^*,\theta_k^*))}{\partial R_k^*(B_k^*,\theta_k^*)}  = (\rho^*  + \mu_k^*) \frac{B_k^*}{R_k^*(B_k^*,\theta_k^*)}.\nonumber
\end{equation}

\noindent First, we analyze $\mu_k^*$. If user $k$ is not assigned any resources, $\mu_k^*>0$ and $B_k^* = 0$. This requires either $\theta_k^* = \infty$ or $p_k = 1$ to satisfy constraint \eqref{eqn:constr2-1}. Substituting $\theta_k^* = \infty$ into \eqref{eqn:qos-guarantee}, we obtain $P_{k}^{\mathrm{th}} = 1$, i.e., the target delay bound violation probability is one. On the other hand, $p_k = {P_{k}^{\mathrm{th}}}^{1/D_{k}^{\mathrm{th}}} = 1$ is equivalent to either $P_{k}^{\mathrm{th}} = 1$ or $D_{k}^{\mathrm{th}} = \infty$. Thus, the optimal bandwidth allocation is always non-zero unless the target delay bound is infinite or the target delay bound violation probability is one. Since this case is not meaningful and impractical, we have $\mu_k^* = 0$ and $B_k^* > 0~\forall k$. Therefore,

\begin{equation}
\frac{\partial Q_k(R_k^*(B_k^*,\theta_k^*))}{\partial R_k^*(B_k^*,\theta_k^*)}  = \rho^* \frac{B_k^*}{R_k^*(B_k^*,\theta_k^*)} .\nonumber
\end{equation}

\noindent Equivalently, we have

\begin{equation}
\frac{\partial Q_k(R_k^*(B_k^*,\theta_k^*))}{\partial R_k^*(B_k^*,\theta_k^*)}\times \frac{R_k^*(B_k^*,\theta_k^*)}{B_k^*}  = \rho^*.\nonumber
\end{equation}

\end{proof}

We refer to the term $R_k^*(B_k^*,\theta_k^*)/B_k^*$  as the \emph{source spectral efficiency} as it represents the number of video bits that can be delivered per channel use subject to the QoS constraint. The interpretation of the result is that the optimal resource allocation is such that the slope of the rate distortion curve multiplied by the source spectral efficiency is the same for all users. Furthermore, that constant determined by $\rho^*$ is computed such that the sum bandwidth constraint is satisfied with equality. Note that as $\rho$ decreases, $B_k^*$ increases for all $k$ and vice versa, thus reducing the problem into a one-dimensional root-finding problem involving solving for $\sum_{k}{B_k^*(\rho)} - B = 0$ to find $\rho$.\\

We make the following observations about the result obtained in Theorem 1:

\begin{enumerate}
\item \textbf{The source spectral efficiency jointly characterizes the channel and the QoS requirement: } The source spectral efficiency $R_k^*(B_k)/B_k$ is independent of the resource allocation $B_k$. It is only a function of the channel statistics, the delay bound, and the delay bound violation probability. Specifically,  $R_k^*(B_k)/B_k = \ln\left(1/P_{k}^{\mathrm{th}}\right)/D_{k}^{\mathrm{th}} T d_k$.
\item \textbf{The resource allocation policy is source spectral efficiency-optimal:} The optimal solution to the quality-maximizing problem is source spectral efficiency-optimal, since $R_k^*(B_k)$ is the maximum rate that could be supported for the given delay bound and violation probability and $R_k^*(B_k)/B_k$ is independent from $B_k$. Thus, $\ln\left(1/P_{k}^{\mathrm{th}}\right)/D_{k}^{\mathrm{th}} T d_k$ is a fundamental measure of the maximum number of video bits per channel use that can be delivered subject to the QoS requirement.
\end{enumerate}

The result in Theorem 1 applies to any fading distribution. In practice, however, solving for the optimal resource allocation and source rates requires numerical computation of $d_k$ by solving $\mathbb{E}_{\boldsymbol{\gamma}}\left\{(1+\gamma_k)^{-\frac{d_k T}{\ln(2)}}\right\} = p_k$. The following corollary simplifies the expression for the special case of a Rayleigh channel.

\begin{Corr1}
For a Rayleigh channel with average SNR $\bar{\gamma}_k$ such that $f_k(\gamma) = \frac{1}{\bar{\gamma}_k}e^{-\gamma/\bar{\gamma}_k}$, the delay constraint reduces to
\begin{equation}
\ln\left(E_{\frac{d_k T}{\ln(2)}}\left(\frac{1}{\bar{\gamma}_k}\right)\right) = \frac{\ln(P_{k}^{\mathrm{th}})}{D_{k}^{\mathrm{th}}} - \frac{1}{\bar{\gamma}_k} + \ln(\bar{\gamma}_k)\nonumber
\end{equation}

\noindent where $E_a(x) = \int_{1}^{\infty}{(e^{-xt}/{t^a})dt}$ is the exponential integral.
\end{Corr1}

\begin{proof}We reduce the delay constraint from Lemma 2 as follows

\begin{eqnarray}
p_k \hspace{-0.2cm}  &=&\hspace{-0.2cm}  \mathbb{E}_{\boldsymbol{\gamma}}\left\{(1+\gamma)^{-\frac{d_k T}{\ln(2)}}\right\} =\frac{1}{\bar{\gamma}_k}\int_{0}^{\infty}{(1+\gamma)^{-\frac{d_k T}{\ln(2)}}e^{-{\gamma}/{\bar{\gamma}_k}}\mathrm{d}\gamma}\nonumber\\
&=& \hspace{-0.2cm}-\frac{1}{\bar{\gamma}_k}e^{1/{\bar{\gamma}_k}} (1+\gamma)^{1-\frac{d_k T}{\ln(2)}} \left.E_{\frac{d_k T}{\ln(2)}}\left(\frac{1+\gamma}{\bar{\gamma}_k}\right)\right]_{0}^{\infty}\nonumber\\
&=& \hspace{-0.2cm}\frac{1}{\bar{\gamma}_k}e^{1/{\bar{\gamma}_k}} E_{\frac{d_k T}{\ln(2)}}\left(\frac{1}{\bar{\gamma}_k}\right)\nonumber
\end{eqnarray}

\noindent where $E_a(x) = \int_{1}^{\infty}{(e^{-xt}/{t^a})dt}$ is the exponential integral. Substituting $p_k = {P_{k}^{\mathrm{th}}}^{1/D_{k}^{\mathrm{th}}}$ and taking the natural logarithm on both sides, the result follows.
\end{proof}

\subsection{Fairness-driven Resource Allocation}

Next, we consider the alternative objective of maximizing the minimum video quality across all served users to provide a notion of fairness in the resource allocation policy. Under this objective, the problem can be formulated as follows

\vspace{-0.3cm}

\begin{eqnarray}
&\hspace{-0.3cm}\max_{\mathbf{R},\mathbf{B}} & \min_{k}{Q_k(R_k)}\label{eqn:obj}\nonumber
\\&\hspace{-0.3cm}\mathrm{s.t.} & \sum_{k=1}^K{B_k} = B\label{eqn:constr1-prob2}\nonumber
\\&\hspace{-0.3cm}& \textrm{Pr}\left\{{D_{k}}>D_{k}^{\mathrm{th}}\right\} \le P_{k}^{\mathrm{th}}~~~\forall~k\label{eqn:constr2-prob2}\nonumber
\\&\hspace{-0.3cm}& B_k \ge 0~~\forall~k\label{eqn:constr3-prob2}\nonumber
\end{eqnarray}

\noindent where $\mathbf{R}$ is the source rate vector and $\mathbf{B}$ is the bandwidth allocation vector.

\begin{Corr2}
The optimal resource allocation vector $\mathbf{B}$ is such that


\vspace{-0.3cm}
\begin{eqnarray}
B_k^* = \frac{T d_k D_{k}^{\mathrm{th}} }{\ln(1/P_{k}^{\mathrm{th}})}Q_k^{-1}(q)\nonumber
\end{eqnarray}

\noindent where $q$ is selected such that $\sum_{k=1}^{K}{T d_k D_{k}^{\mathrm{th}}Q_k^{-1}(q)/\ln(1/P_{k}^{\mathrm{th}})} = B$.
\end{Corr2}

\begin{proof}
Given any resource allocation $\{B_k\}_{k=1}^K$, let $k' = \mathrm{argmin}_{k}{Q_k(R_k^*(B_k)}$. Note that, from Lemma 3, for any resource allocation $\{B_k\}_{k=1}^K$, a feasible solution ($\mathbf{R}^*,\mathbf{B}^*$) always exists if $Q_k(R_k)$ is defined for every $R_k\ge 0$. Thus, $k'$ is well defined. By increasing $B_{k'}$, the objective function $\min_{k}{Q_k(R_k)} = Q_{k'}(R_{k'})$ keeps improving until $\mathrm{argmin}_{k}{Q_k(R_k^*(B_k)}\ne k'$. Thus, it follows that the optimal solution requires $Q_k(R_k^*(B_k^*) = Q_j(R_j^*(B_j^*,\theta_j^*)) = q \forall j,k$. Furthermore, given that $Q_k(R_k)$ is defined for every $R_k\ge 0$, a small enough $q$ always exists. Given any target video quality $q$ for each user, using Lemma 1 and Lemma 2, the corresponding resource allocation for user $k$ is

\begin{eqnarray}
B_k  = \frac{d_{k}}{\theta_{k}} = \frac{T d_k D_{k}^{\mathrm{th}} }{\ln(1/P_{k}^{\mathrm{th}})}Q_k^{-1}(q)\nonumber
\end{eqnarray}

\noindent Thus, the required sum bandwidth for all users to a video quality $q$ for each user is

\begin{eqnarray}
\sum_{k=1}^{K}{B_k}  = T\sum_{k=1}^{K}{\frac{d_k D_{k}^{\mathrm{th}}Q_k^{-1}(q)}{\ln(1/P_{k}^{\mathrm{th}})}}\nonumber
\end{eqnarray}

\noindent and $B_k^*$ can be found by numerically solving for $q$ such that $\sum_{k=1}^{K}{B_k} = B$.
\end{proof}

\subsection{Practical Considerations}

In practice, the number of operating points on the rate-distortion curve is finite and determined by the different representations of the video sequence available at the server. Let $(R_k^{\textrm{min}},Q_k^{\textrm{min}})$ correspond to the operating point on user $k$ rate distortion curve that provides the lowest available rate. If $R_k^*(B_k^*,\theta_k^*) < R_k^{\textrm{min}}$, then the solution corresponding to user $k$ cannot be realized. Thus, practically, user $k$ cannot use any of the available video descriptions and still meet the target delay constraint.

\section{Maximal User Subset Scheduling and User Admission}\label{sec:scheduling}

In this section, we solve the problem of selecting the largest subset of users to serve such that each user in the subset can meet their delay constraint. We solve the problem under sum quality-maximizing resource allocation as well as fairness-based resource allocation. Afterwards, we present a user admission policy that determines the criterion for a new user to be admitted into the system given the delay constraint and the delay constraints for the existing users so that the delay constraint of each user is not compromised.

Theorem 2 summarizes the main result of the section. It presents the optimal scheduling policy that selects a maximal subset of users such that all user meet their statistical delay constraint. To summarize, the optimal scheduling policy can be obtained in polynomial time in the number of users and it involves computing the minimum bandwidth required by each user to support their QoS requirement, using it as a sorting criterion, and scheduling the first sorted users such that the sum of their minimum bandwidth requirement does not exceed the total bandwidth. Under the fairness constraint, a similar solution is obtained with the major difference that the sorting criterion is the video quality corresponding to the minimum rate representation of the video sequence.

\begin{Thm2}
\textbf{Maximal User Subset Scheduling:} Under maximum sum quality resource allocation, the scheduling policy to maximize the number of users that can meet their QoS requirement is as follows. Define

\begin{equation}
B_k^{\min} = \frac{D_{k}^{\mathrm{th}} d_k T}{\ln(1/P_{k}^{\mathrm{th}})} R_k^{\min}
\end{equation}

\noindent and let $\ell(i)$ be the sorting operation in increasing order on $B_k^{\min}$ so that ${B_{\ell(i)}^{\min}}$ is the $i^\mathrm{th}$ sorted element $\forall i=1,\cdots,K$, then the maximum number of users supported $N^*$ is

\begin{equation}
N^* = \mathrm{argmax}_N \sum_{i=1}^{N}{B_{\ell(i)}^{\min}} ~\mathrm{s.t.}~ \sum_{i=1}^{N}{B_{\ell(i)}^{\min}}\le B\nonumber
\end{equation}

\noindent and the corresponding scheduled users are $\ell(1),\cdots,\ell(N^*)$.

Under fairness-based resource allocation, the scheduling policy to maximize the number of users that can meet their QoS requirement is as follows. Define $q_k^{\min} = Q_k(R_k^{\min})$ and let $m(i)$ be the sorting operation in increasing order on $q_k^{\min}$ so that ${q_{m(i)}^{\min}}$ is the $i^{th}$ sorted element $\forall i=1,\cdots,K$, then the maximum number of users supported $N^*$ is

\begin{equation}
N^* = \mathrm{argmax}_N \sum_{i=1}^{N}{B_{m(i)}(q_{m(N)}^{\min})} ~\textrm{s.t.}~ \sum_{i=1}^{N}{B_{m(i)}(q_{m(N)}^{\min})}\le B\nonumber
\end{equation}

\noindent and the corresponding scheduled users are $m(1),\cdots,m(N^*)$.
\end{Thm2}

\begin{proof}
If user $k$ is scheduled, the minimum bandwidth requirement is such that the user's delay constraint can be met with the minimum source rate. As the source rate increases, a larger bandwidth is required to maintain the same delay constraint. Thus, the minimum bandwidth ${B_{k}^{\min}}$ required to meet the statistical delay bound is

\begin{equation}
B_k^{\min} = B_k^*(R_k^{\min}) = \frac{D_{k}^{\mathrm{th}} d_k T}{\ln(1/P_{k}^{\mathrm{th}})} R_k^{\min}.\nonumber
\end{equation}

\noindent Given the minimum bandwidth required per user, the problem of finding a maximal user subset reduces to a special case of the Knapsack problem \cite{kellerer2004knapsack} where all items (users) have equal value $v_k = 1$ and weights (bandwidths) $w_k = B_k^{\min}$. Since $v_k = 1~\forall k$, the exact solution corresponds to the $N^*$ users with the smallest $w_k$ such that $\sum{w_k}<B$. More formally, define $\ell(i)$ to be the sorting operating in increasing order so that ${B_{\ell(i)}^{\min}}$ is the $i^{th}$ sorted element $\forall i=1,\cdots,K$, then the maximum number of users supported $N^*$ is

\begin{equation}
N^* = \mathrm{argmax}_N \sum_{i=1}^{N}{B_{\ell(i)}^{\min}} ~\mathrm{s.t.}~ \sum_{i=1}^{N}{B_{\ell(i)}^{\min}}\le B\nonumber
\end{equation}

\noindent and it follows that the corresponding scheduled users are $\ell(1),\cdots,\ell(N^*)$.

Under the fairness-based resource allocation, define $m(i)$ be the sorting operating in increasing order so that ${q_{m(i)}^{\min}}$ is the $i^{th}$ sorted element $\forall i=1,\cdots,K$. Corollary 2 states that, under the fairness-based policy, all users should achieve the same target video quality $\bar{q}$. Defining $q_k^{\min} = Q_k(R_k^{\min})$, it follows that if $q_k^{\min}>\bar{q}$, user $k$ cannot be served. Thus, user $m(i)$ can only be served if $m(1),\cdots,m(i-1)$ are also served. This reduces the number of possible solutions to $K+1$. The first possibility is serving no users, the second is serving user $m(1)$, all the way to option $K+1$ which requires serving users $m(1),\cdots,m(K)$. Furthermore, if users $m(1),\cdots,m(N)$ are served, the minimum target quality is $q_{m(N)}^{\min}$ since all users are required to maintain the same quality and any $\bar{q}<q_{m(N)}^{\min}$ is not feasible for at least one user. Thus, the maximum number of users supported $N^*$ can be found as follows

\begin{equation}
N^* = \mathrm{argmax}_N \hspace{-0.1cm}\sum_{i=1}^{N}{B_{m(i)}(q_{m(N)}^{\min})} ~\mathrm{s.t.}~\hspace{-0.2cm} \sum_{i=1}^{N}{B_{m(i)}(q_{m(N)}^{\min})}\le B\nonumber
\end{equation}

\noindent and it follows that the corresponding scheduled users are $m(1),\cdots,m(N^*)$.
\end{proof}

Next, we contrast the scheduling solution under maximum sum quality resource allocation with that under fairness-based resource allocation. First, with maximum sum quality resource allocation, the system can always support at least the same number of users as with the fairness-based policy. This can be seen in the following inequalities

\begin{equation}
B_{\ell(i)}^{\min} \le B_{m(i)}^{\min}\nonumber \le B_{m(i)}(q_{m(N)}^{\min})\nonumber
\end{equation}

\noindent where the first inequality follows since $\ell(i)$ is the $i^{\mathrm{th}}$ sorted element according to the $B_{k}^{\min}$ criterion and the second inequality follows by the definition of $B_{k}^{\min}$. Furthermore, no scheduling policy can serve more users than that under maximum sum quality resource allocation since it is based on each served user operating at the minimum bandwidth required to maintain the QoS requirement. Finally, the fairness-based policy and the quality-maximizing policy support the same number of users if $Q_k(R) = Q(R) \forall k$ and $R_k^{\min} = R^{\min} \forall k$, that is, all users have the same rate-distortion behavior and minimum rate representations. While this is typically not the case in practice, it provides the following useful intuition: Achieving fairness is least costly when the rate-distortion behavior of the users in the network is similar such that all users achieve similar incremental gains in quality when increasing their source rate.

We next consider the user admission problem, whereby a set of users are already using the wireless system resources for video streaming, and a new user enters the system. For this scenario, we derive the user admission criterion that ensures that (1) the admitted user meets the target QoS requirement, and (2) the current users' QoS requirements' are not jeopardized by the admitted user. This user admission policy is presented in Theorem 3.

\begin{Thm3}
\textbf{User Admission Criteria:} Under the quality-maximizing policy, a new user with QoS requirements $\{D_{K+1}^{\mathrm{th}},P_{K+1}^{\mathrm{th}}\}$ can be admitted into the system with $K$ existing users operating at $\{R_k\}^{K}_{k=1}$ if

\begin{eqnarray}
\frac{R_k^{\textrm{min}}}{R_k}&\le& 1-\frac{R_{K+1}^*}{B} \times\frac{d_{K+1}T D_{K+1}^{\mathrm{th}}}{\ln(1/P_{K+1}^{\mathrm{th}})} ~\forall k=1,\cdots,K\nonumber\\
\frac{R_{K+1}^{\textrm{min}}}{R_{K+1}^*}&\le& 1.\nonumber
\end{eqnarray}

\noindent Under the quality fairness policy where each of the $K$ user operates with quality $q^\textrm{init} = Q_i(R_i)$, user $K+1$ with QoS requirements $\{D_{K+1}^{\mathrm{th}},P_{K+1}^{\mathrm{th}}\}$ can be admitted if

\begin{eqnarray}
&&\hspace{-0.5cm}\frac{R_i^{\textrm{min}}}{R_i}\le 1-\frac{Q_{K+1}^{-1}(q^\textrm{final})}{B} \times\frac{d_{K+1}T D_{K+1}^{\mathrm{th}}}{\ln(1/P_{K+1}^{\mathrm{th}})} ~\forall i=1,\cdots,K\nonumber\\
&&\hspace{-0.5cm}\frac{R_{K+1}^{\textrm{min}}}{Q_{K+1}^{-1}(q^\textrm{final})}\le 1\nonumber
\end{eqnarray}

\noindent where $q^\textrm{final}$ is computed such that $\sum_{k=1}^{K+1}{B_k^*} = B$.
\end{Thm3}

\begin{proof}
For user $K$ to be admitted, two conditions need to be satisfied. First, the other users need to be able to maintain their own QoS constraint. Second, the admitted user needs to be allocated enough resources to meet the target QoS requirement. If admitted, user $K+1$ gets bandwidth

\begin{equation}
B_{K+1}^*(R_{K+1}) = \frac{d_{K+1}}{\theta_{K+1}^*} = R_{K+1} \frac{d_{K+1} T D_{K+1}^{\mathrm{th}}}{\ln(1/P_{K+1}^{\mathrm{th}})}.\nonumber
\end{equation}

\noindent The new resource allocation for the rest of the users is obtained by allocating an equivalent total bandwidth of $B-B_{K+1}^*$. The source rate scales linearly in the bandwidth according to Theorem 1, i.e., $R_k(B_k)/B_k$ is independent of $B_k$ for each user $k$. Thus, since the old source rate for user $k$ is $R_k = B\times(R_k/B)$, admitting user $K+1$ reduces user $k$'s source rate to $R_k^{\mathrm{new}} = (B - B_{K+1}^*)\times(R_k/B)$. Thus, the user admission condition reduces to

\begin{eqnarray}
R_k^{\min} \le R_k^{\mathrm{new}} &=& \frac{R_k(B - B_{K+1}^*)}{B}
\nonumber\\&=& R_k\left( 1 - R_{K+1} \frac{d_{K+1} T D_{K+1}^{\mathrm{th}}}{B\ln(1/P_{K+1}^{\mathrm{th}})}\right).\nonumber
\end{eqnarray}

\noindent Thus, the condition can be written equivalently as

\begin{eqnarray}\label{eqn:user-admission-criterion}
\frac{R_k^{\min}}{R_k} \le 1 - \frac{R_{K+1}^*}{B} \times\frac{d_{K+1} T D_{K+1}^{\mathrm{th}}}{\ln(1/P_{K+1}^{\mathrm{th}})}.
\end{eqnarray}

\noindent If the condition in \eqref{eqn:user-admission-criterion} is satisfied for each user $k$, the QoS constraint is maintained after user $K+1$ is admitted. In addition, for user $K+1$ to satisfy their own QoS requirement, the maximum source rate required should exceed the minimum rate representation $R_{K+1}^{\min}$, i.e., $R_{K+1}^{\min}\le R_{K+1}^*$. Under the fairness-based policy, the same analysis follows with the exception that $R_{K+1}^* = Q_{K+1}^{-1}(q^\textrm{final})$ such that, after user admission, all users have equal video quality $q^\textrm{final}$ and $q^\textrm{final}$ is selected such that $\sum_{k=1}^{K+1}{B_k}=  B$.
\end{proof}

\begin{figure*}[tbp!]
    \centering
    \subfigure[Average SNR $\bar{\gamma} = 0$ dB.]{\includegraphics[height=6.1cm]{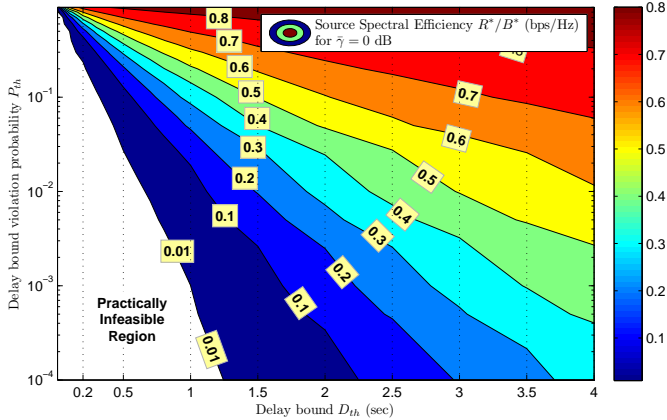}}
    \subfigure[Average SNR $\bar{\gamma} = 20$ dB.]{\includegraphics[height=5.9cm]{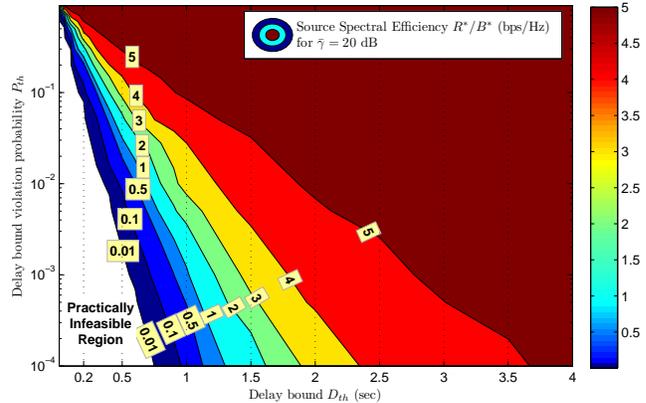}}
    \caption{\label{fig:SE1} Contour plot of source spectral efficiency, the maximum source bits per channel use, for a Rayleigh channel vs. delay bound $D^{\mathrm{th}}$ and delay bound violation probability $P^{\mathrm{th}}$. In the lower left regime, the steep decline puts a practical limitation on the feasible QoS operating points.}
\end{figure*}

\section{Joint Scheduling and Resource Allocation Algorithm}\label{sec:Alg}

In this section, we present the Algorithm that jointly selects a maximal user subset from a set of candidate video users with delay QoS constraints. Among the scheduled users, resources are allocated either to maximize the sum video quality or minimum video quality. The Algorithmic description for the sum quality maximizing policy is provided in Algorithms 1.

First, we use the delay constraint ${D_{k}^{\mathrm{th}}}$ and the delay constraint violation probability ${P_{k}^{\mathrm{th}}}$ to compute $p_k$ the bandwidth-QoS exponent product $d_k$. Next, we compute the minimum bandwidth requirement $B_k^{\min}$ for each user to meet their QoS constraint and the corresponding video quality $q_k^{\min}$. Based on the channel conditions and the QoS requirement, the maximum source spectral efficiency $R_k^*(B_k)/B_k$ is computed for each user. The next step involves scheduling a maximal user subset based on Theorem 2 depending on the selected utility function. After a subset of users is selected, the optimal bandwidth allocation $B_k^*$ is obtained using Theorem 1.

%

\vspace{0.3cm}

\begin{boxedalgorithmic}{1}{Quality-maximizing Joint Scheduling and Resource Allocation.}

\STATE Given $K$ users with delay QoS requirements $\{D_{k}^{\mathrm{th}},P_{k}^{\mathrm{th}}\}_{k=1}^{K}$, fading distributions $f_k(\gamma)$, and total bandwidth $B$.

\STATE \underline{\textbf{\emph{Step 1. Compute QoS-related Metrics}}}

\FOR{$k = 1 \to K$}

\STATE
\begin{enumerate}
\item Find $p_k = {P_{k}^{\mathrm{th}}}^{1/D_{k}^{\mathrm{th}}}~\forall k$.

\item  Given $f_k(\gamma)$, find the bandwidth-QoS exponent product $d_k = B_k^* \theta_k^* ~\forall k$ by solving $\mathbb{E}_{\boldsymbol{\gamma}}\left\{(1+\gamma_k)^{-\frac{d_k T}{\ln(2)}}\right\} = p_k$.

\item  Compute minimum bandwidth requirement per user $B_k^{\min} = \frac{D_{k}^{\mathrm{th}} d_k T}{\ln(1/P_{k}^{\mathrm{th}})} R_k^{\min}$ and minimum quality $q_k^{\min} = Q_k(R_k^{\min})$.

\item  Compute the maximum source spectral efficiency $R_k^*(B_k)/B_k = \ln\left(1/P_{k}^{\mathrm{th}}\right)/D_{k}^{\mathrm{th}} T d_k$.
\end{enumerate}

\ENDFOR

\STATE \underline{\textbf{\emph{Step 2. Select Maximal User Subset }}}

\STATE \begin{enumerate}
\item Sort ${B_{k}^{\min}}$ with operator $\ell(i)$ such that ${B_{\ell(i)}^{\min}}$ is the $i^{th}$ sorted element $\forall i=1,\cdots,K$

\item Compute maximum number of users $N^* = \textrm{argmax}_N \sum_{i=1}^{N}{B_{\ell(i)}^{\min}} ~\textrm{s.t.}~ \sum_{i=1}^{N}{B_{\ell(i)}^{\min}}\le B$

\item Schedule users $\ell(1),\cdots,\ell(N^*)$.
\end{enumerate}

\STATE \underline{\textbf{\emph{Step 3. Allocate Resources}}}

\STATE
\begin{enumerate}
\item Set $\rho = \rho_{\mathrm{initial}}$

\item Solve $\frac{\partial Q_k(R_k^*(B_k^*,\theta_k^*))}{\partial R_k^*(B_k^*,\theta_k^*)}  = \frac{T d_k^* (\rho^*)}{\ln\left(p_k\right)}$ for $R_k^*~\forall k = \ell(1),\cdots,\ell(N^*)$

\item Solve $\theta_k^* = -\ln(p_k)/(T R_k^*)~\forall k = \ell(1),\cdots,\ell(N^*)$

\item Solve $B_k^* = d_k/\theta_k^*~\forall k = \ell(1),\cdots,\ell(N^*)$\\

\item If $\sum{B_k^*}<B$, decrease $\rho$ and repeat 2,3,4, otherwise increase $\rho$ and repeat 2,3,4 until $|\sum{B_k^*}-B|<\epsilon$.

\end{enumerate}

\end{boxedalgorithmic}\vspace{1cm}

\section{Results}\label{sec:results}

In this section, we present results and analysis to demonstrate the performance of the proposed scheduling and resource allocation algorithms for real-time video transmission. First, we present analysis of the source spectral efficiency under different QoS requirements and channel conditions. Next, we analyze the resource allocation for the two user case with hybrid QoS requirements. Finally, we present the general case of multiple users with joint scheduling and resource allocation.

\subsection{Source Spectral Efficiency under Delay Constraints}

Figure \ref{fig:SE1} shows a contour plot of the maximum source spectral efficiency supported by a Rayleigh channel as a function of the delay bound and delay bound violation probability. Note, from Theorem 1, that the maximum source spectral efficiency can be expressed as $\ln\left(1/P_{k}^{\mathrm{th}}\right)/D_{k}^{\mathrm{th}} T d_k$ and is achieved by the proposed rate adaptation algorithm. The main observation from the plot is that there exists a boundary beyond which the source spectral efficiency declines very rapidly, making practical realization of the QoS constraint impossible since it would require either using a very large bandwidth or operating at a very small source rate. For example, to achieve $D_{k}^{\mathrm{th}} = 0.5$ sec and $P_{k}^{\mathrm{th}} = 10^{-3}$, we require a bandwidth 1000 times the source rate even if the average SNR is $\bar{\gamma} = 20$ dB. This provides a practical insight into the range of feasible QoS constraints.


\begin{figure}[tbp!]
    \centering
\includegraphics[width=9.5cm]{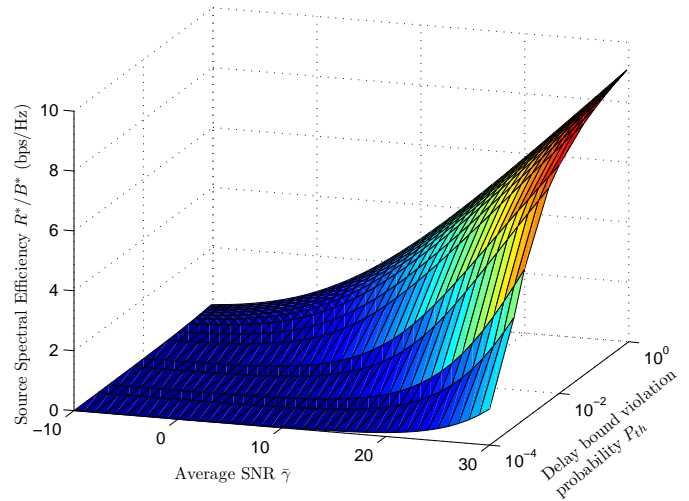}
    \caption{\label{fig:SE2} Source spectral efficiency for a Rayleigh channel vs. average SNR $\bar{\gamma}$ and target delay bound violation probability $P^{\mathrm{th}}$ for a target delay bound $D^{\mathrm{th}} = 0.2$ sec.}
\end{figure}

Figure \ref{fig:SE2} shows the achievable source spectral efficiency over a range of SNRs for $D_{k}^{\mathrm{th}} = 0.2$ sec. It can be seen that the advantage of good channel conditions is minimal under very stringent QoS requirements whereas relaxing the target delay violation probability enables supporting high source rates even at poor channel conditions. We note that these results are considered as an extreme case since they consider a Rayleigh channel and uncorrelated fading instances. With an uncorrelated channel, the randomness in the service rate is largest, thus exacerbating the problem of maintaining a certain delay constraint.

\subsection{Resource Allocation under Hybrid Delay QoS Requirements}

\begin{figure*}[tbp!]
    \centering
    \subfigure[Service regions.]{\includegraphics[height=6.5cm]{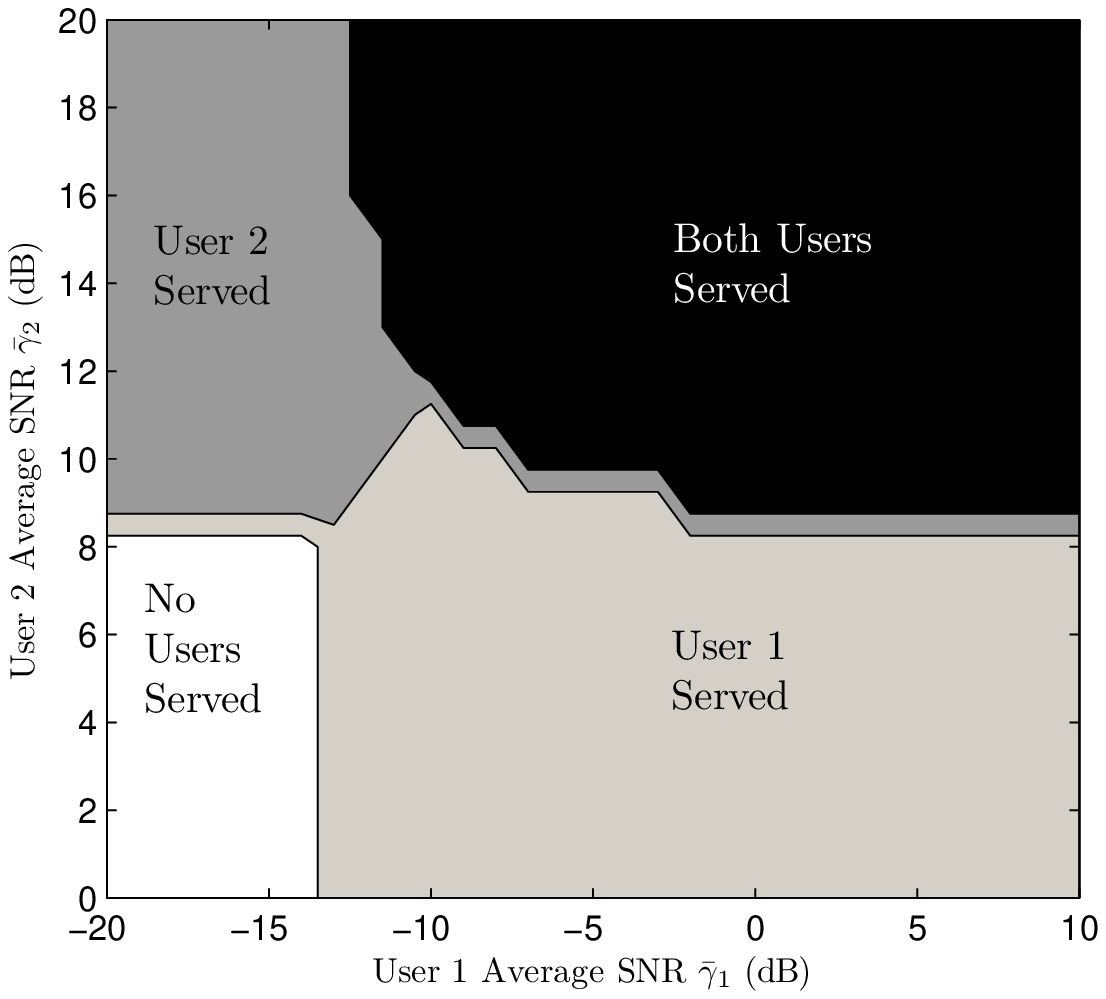}}
    \subfigure[Corresponding average quality (STRRED) per user.]{\includegraphics[height=6.5cm]{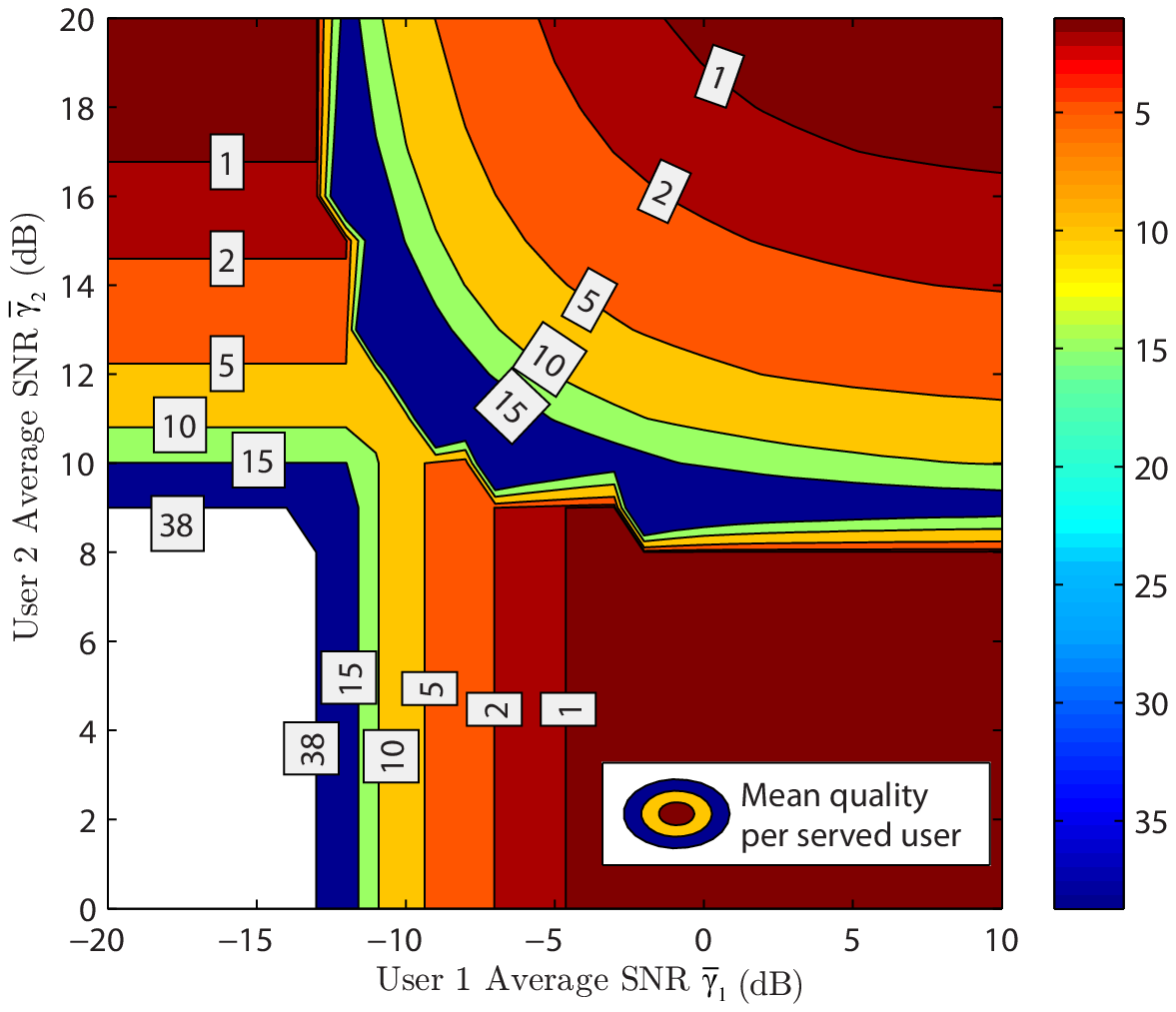}}
    \caption{\label{fig:2user} Two user case study with unequal QoS requirements, that is, $D_{1}^{\mathrm{th}} = 2$ sec, $D_{2}^{\mathrm{th}} = 0.3$ sec, and $P_{1}^{\mathrm{th}} = P_{2}^{\mathrm{th}} = 0.1$. The minimum source rate for a user to be served is $R_1^{\min} = R_2^{\min} = 185$ Kbps, and $B = 5$ MHz. The significant difference in SNR to achieve the same video quality shows that scheduling based only on channel conditions is highly suboptimal. Note that lower STRRED values corresponds to higher video quality. }
\end{figure*}

Next, we consider two users with different QoS requirements sharing a wireless channel. User 1 has a delay constraint $D_{1}^{\mathrm{th}} = 2$ sec corresponding to a typical live video streaming application and user 2 has a delay constraint $D_{2}^{\mathrm{th}} = 0.3$ sec corresponding to a typical interactive video conferencing application. The target delay bound violation probability is $0.1$ for both users. To understand the effect of unequal delay requirements in isolation, we consider the case where both users have the same rate distortion characteristics. Specifically, we use the Foreman video sequence \cite{videoSequences} encoded with H.264/AVC. The GoP structure is $IBPBP\cdots$ and the GoP duration is 16 frames. The MB size is $16\times16$ and we use the CIF resolution of $352 \times 288$.
The different rate-distortion operating points are obtained by modifying the quantization parameters of the discrete cosine transform (DCT) coefficients. We measure video quality using the spatio-temporal reduced reference entropic differencing (STRRED) index \cite{ST-RRED}. It is a reduced reference (RR) video quality metric that uses combined spatial and temporal information differences and correlates quite well with subjective quality. Further, it requires minimal exchange of side information from the video server, making it suitable for resource allocation at the network edge. We specifically use $\mathrm{STRRED}_4^{M_4}$ corresponding to the vertically oriented subband which is shown in \cite{ST-RRED} to perform best. Note that the higher values of STRRED correspond to lower video quality. Thus, our proposed algorithm uses the rate-quality function $Q_k(R_k) = - \mathrm{STRRED}_4^{M_4}(R_k)$.

The minimum rate representation is $R^{\min} = 185$ Kbps and it provides a corresponding STRRED of $38$. The total system bandwidth is 5 MHz. Figure \ref{fig:2user}a shows the range of SNRs over which: (1) None of the users can be served, (2) only user 1 can be served, (3) only user 2 can be served, and (4) both users can be served simultaneously. The main observation is that the SNR range is very different for the two users. User 1 can be served with as low as -$14$ dB while user 2 requires at least 8 dB to satisfy their QoS requirement. This shows that Max SNR-based scheduling is highly suboptimal for video users with different QoS requirements and QoS-aware scheduling achieves significant gains. For instance, for $-14\le\bar{\gamma}_2\le8$ and $\bar{\gamma}_1<\bar{\gamma}_2$, a Max SNR scheduler always favors user 2, although it is clear that in this entire operating region, user 2 cannot meet the QoS requirement while user 1 can. Figure \ref{fig:2user}b shows a contour plot of the corresponding average video quality per user at each operating point.

\subsection{Multiuser Scheduling and Resource Allocation}

\begin{figure*}[tbp!]
    \centering
    \subfigure[Average number of users supported]{\includegraphics[width=8.1cm]{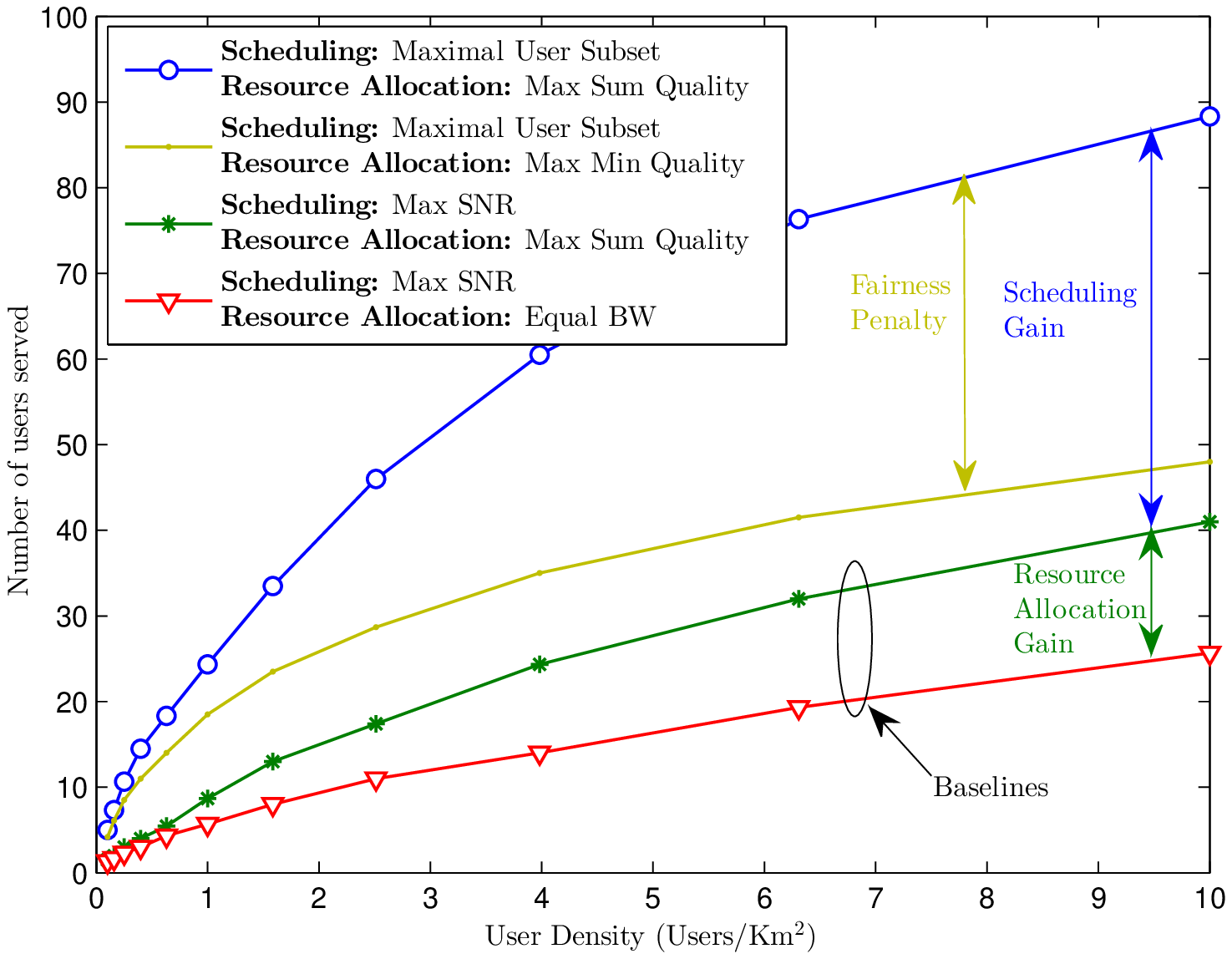}}
    \subfigure[Standard deviation of DMOS-mapped video quality]{\includegraphics[width=8.1cm]{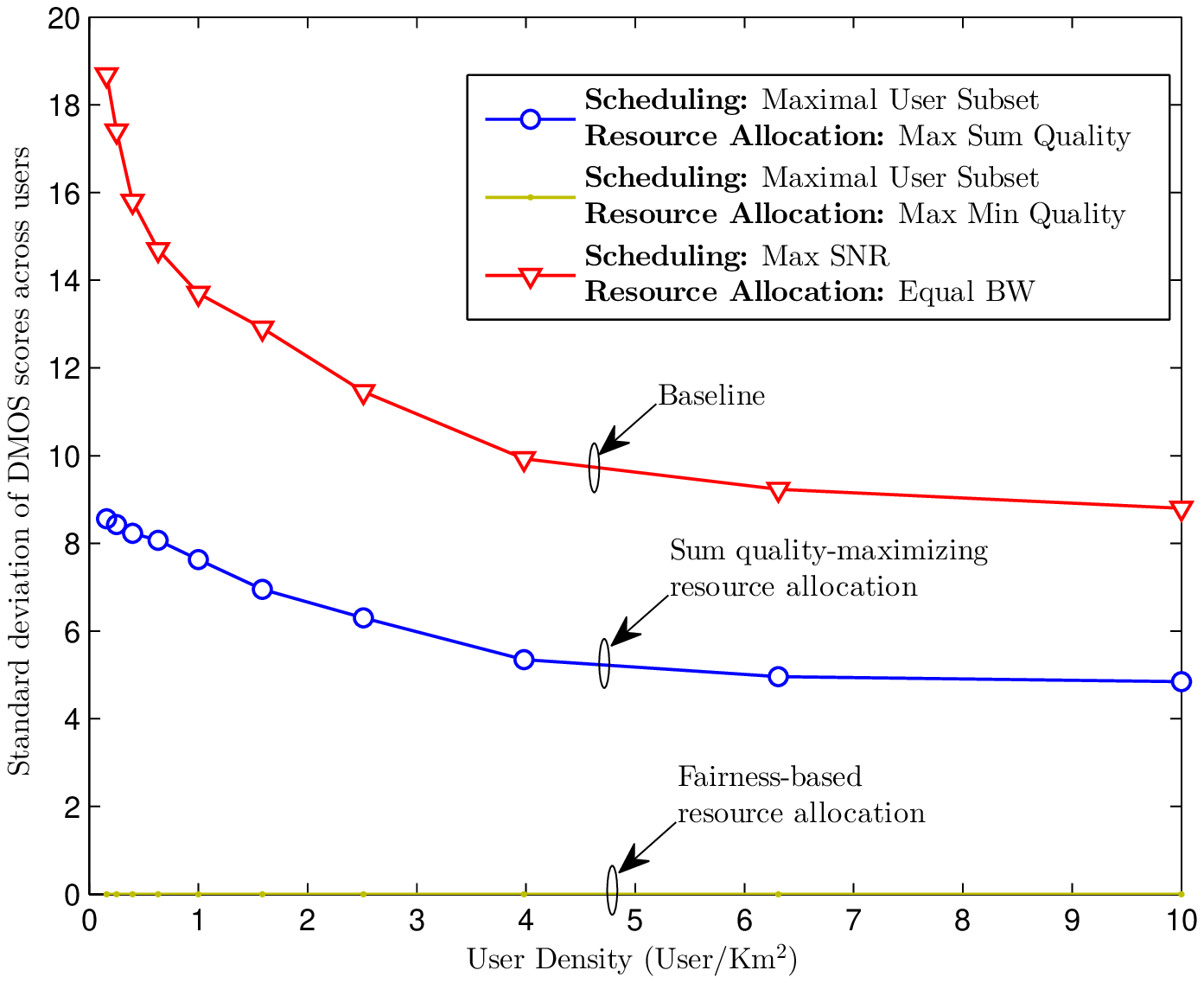}}
    \caption{\label{fig:multiuser} Analysis of proposed scheduling and resource allocation algorithms in comparison to the baselines vs. user density for $P_t=30$ dBm and $B$ = 20 MHz. The scheduling baseline is QoS-aware Max SNR, that is, the maximum number of the highest SNR users that can meet their delay constraint are scheduled. In contrast to Max sum quality and fairness-based resource allocation, the resource allocation baseline shares the resources equally among scheduled users.}
\end{figure*}

Next, we consider the general case where there is a large number of users to be served. We apply the scheduling solution derived in Theorem 2 to select a user subset followed by the corresponding resource allocation policy. We consider a single cell setup where the users are distributed according to a poisson point process (PPP) in the cell. Half the video users have a delay constraint $D_{1}^{\mathrm{th}} = 2$ sec corresponding to live video users and the other half have a delay constraint $D_{2}^{\mathrm{th}} = 0.3$ sec corresponding to a typical two-way video conferencing user. The target delay bound violation probability is $0.1$ for both sets of users and the total system bandwidth is 20 MHz. We use three video sequences \cite{videoSequences} encoded with H.264/AVC. The GoP structure is $IBPBP\cdots$ and the GoP duration is 16 frames. The MB size is $16\times16$ and we use the CIF resolution of $352 \times 288$. The minimum rate representation of the three different video are $R_k^{\min} = 160$, $185$, and $200$ Kbps. For accurate channel modeling, we derive the average SNRs $\bar{\gamma}_{k}$ from the distances $d_{k}$ using the following relation

\vspace{-0.3cm}
\begin{equation}
\bar{\gamma}_{k} = 10\log_{10}(P_{t}) - K_{\textrm{dB}} - 10\delta \log_{10}(d_{k}) - 10\log(N_0 B_k)\nonumber
\end{equation}

\noindent where $K_{\textrm{dB}}$ is the pathloss constant, $\delta$ is the pathloss exponent, and $N_0$ is the power spectral density of AWGN noise. In our simulations, $ K_{\textrm{dB}} = 21.36 ~\textrm{dB}$, $\delta = 3.52$, and $N_0 = 4\times 10^{-21}$~W/Hz.

Figure \ref{fig:multiuser}a shows the number of users supported using the proposed maximal user subset scheduling algorithm along with (a) sum quality-maximizing resource allocation, and (b) fairness-based resource allocation. As we proved in Theorem 2, the maximal user subset scheduling algorithm with sum quality-maximizing resource allocation outperforms any other scheduling/resource allocation combination under the same delay constraints. To distinguish the resource allocation and scheduling gains, we consider two baselines. In both baselines, scheduling is based on QoS-aware Max SNR, that is, the maximum number of the highest SNR users that can be supported such that they can meet their delay constraint is scheduled. In the first baseline, the available bandwidth is divided equally among the scheduled users. In the second baseline, the bandwidth is allocated according to the sum quality-maximizing resource allocation. Thus, the difference between the baselines is the resource allocation gain and the difference between the second baseline and the proposed algorithm is the scheduling gain. We observe that significant gains are achieved in terms of the total number of users supported in the system. The resource allocation gain corresponds to 1.6x increase in capacity due to the better partitioning of the wireless system resources. The scheduling gain corresponds to 2.2x-3.5x increase in capacity. Furthermore, under the fairness constraint, the number of users supported drops because forcing users with hybrid QoS requirements and different channel conditions to operate at the same video quality increases the resource consumption, particularly for cell edge users. Still, it outperforms both baselines.

Figure \ref{fig:multiuser}b shows the standard deviation of the video quality across users vs. user density under the sum quality-maximizing policy. To obtain a meaningful interpretation of the standard deviation, we first map the STRRED quality scores to differential mean opinion score (DMOS) which is linear in user judgments and ranges from 0 to 100. For that purpose, we use the fit in \cite{ST-RRED}. We deduce from Figure \ref{fig:multiuser}b that the deviation of quality across users is reasonable. Thus, maximizing sum quality comes at a little cost to cell edge users' performance and maintaining \emph{complete fairness} by maximizing the minimum video quality in the cell is too costly to the capacity.

\begin{figure*}[tbp!]
    \centering
    \subfigure[Max subset scheduler and max sum quality allocation.]{\includegraphics[width=8cm]{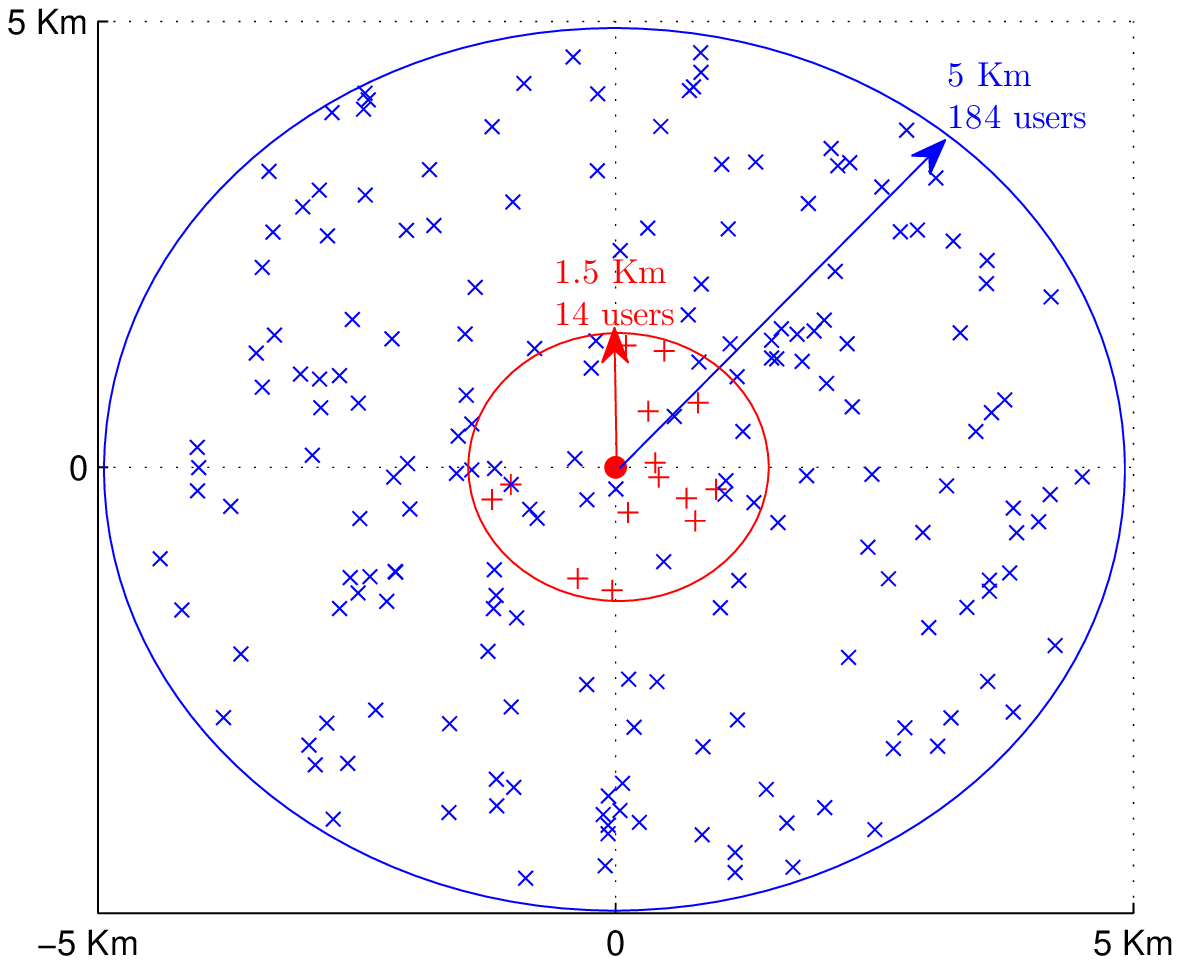}}
    \subfigure[Max SNR scheduler and equal bandwidth allocation.]{\includegraphics[width=8cm]{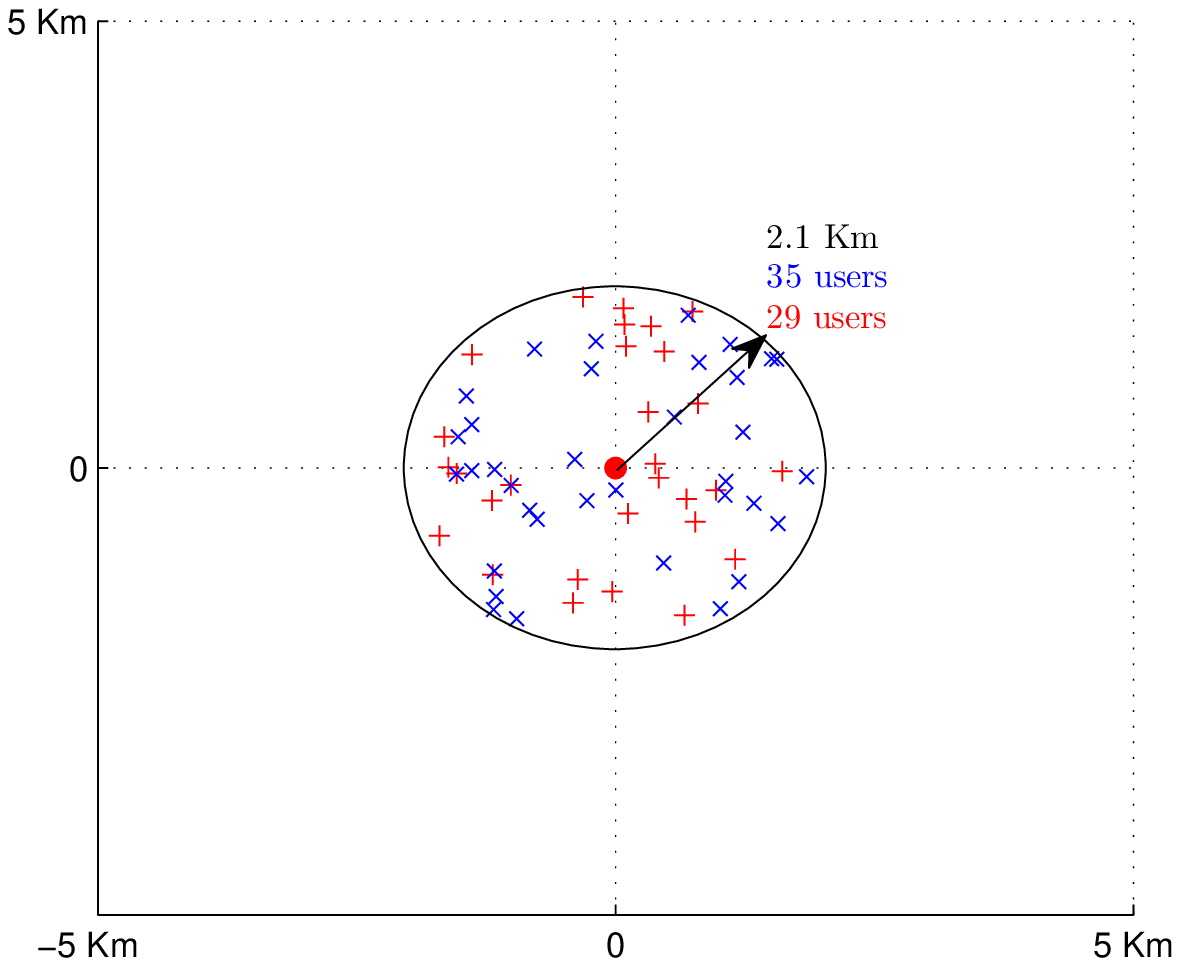}}
    \caption{\label{fig:coverage} Users served by the proposed and baseline scheduling and resource allocation algorithms for $P_t = 25$ W. Red markers correspond to video conferencing sessions and blue markers correspond to live streaming users. The proposed maximal user subset scheduling drops a small fraction of users that require excessively large amount of resources to meet their QoS requirement such that the total number of users supported is maximized.}
\end{figure*}

To understand how the significant capacity gains are achieved, Figure \ref{fig:coverage} shows the users covered by the proposed scheduling and resource allocation algorithm in comparison to the first baseline at $P_t = 25 $ W $= 44$ dBm. The maximal user subset scheduling algorithm supports a total of 198 users corresponding to 184 live streaming sessions and 14 video conferencing sessions. In contrast, Max SNR scheduling supports only 64 users corresponding to 35 live streaming sessions and 29 video conferencing sessions. Thus, in essence, the gains stem from dropping a small fraction of users that require excessively large amount of resources to meet their QoS requirement such that the total number of users supported is maximized. Another way to visualize the gains is through the coverage radius. With the baseline, the coverage radius is 2.1 Km. Under the proposed algorithm, the coverage radius for video conferencing users is reduced to 1.5 Km such that live streaming users can be served up to a 5 Km radius.


\section{Conclusion}\label{sec:conc}

In this paper, we used the concept of effective capacity to provide a framework for statistical delay provisioning for multiple users sharing a wireless network. Sum quality-maximizing resource allocation policies as well as fairness-based policies were derived. Furthermore, maximal user subset scheduling was proposed to maximize the number of scheduled users that can meet their QoS requirement. Significant gains in capacity, measured in terms of number of users supported in the system, are achieved due to QoS-aware scheduling and resource allocation.


%
%
%
%
%
%
%
%

\vspace{-0.1cm}
\bibliographystyle{ieeetr}
\bibliography{references}

\end{document}